\PassOptionsToPackage{numbers}{natbib}
\documentclass{article}
\usepackage[main, final]{neurips_2026}

\usepackage[utf8]{inputenc} 
\usepackage[T1]{fontenc}    
\usepackage{hyperref}       
\usepackage{url}            
\usepackage{booktabs}       
\usepackage{amsfonts}       
\usepackage{nicefrac}       
\usepackage{microtype}      
\usepackage{xcolor}         
\usepackage{amsmath} 
\usepackage{graphicx}
\usepackage{amssymb}
\usepackage{pifont}
\usepackage{wrapfig}
\usepackage{caption}
\usepackage{chngcntr}
\usepackage[table]{xcolor}

\title{HELICS: Biobank-scale Conditional Synthetic Genome Generation via Latent Flow Matching} 

\author{
  Ahmad Abdel-Azim \\
  Department of Biostatistics\\
  Harvard University\\
  Boston, MA 02135 \\
  \texttt{ahmad\_abdelazim@g.harvard.edu} \\
  \And
  Xihong Lin \\
  Department of Statistics, Biostatistics\\
  Harvard University\\
  Boston, MA 02135 \\
  \texttt{xlin@hsph.harvard.edu}
}

\begin{document}

\maketitle

\begin{abstract}
    Individual-level genotype data underpin genetic discovery, disease-risk modeling, and therapeutic development, yet many critical settings remain data-limited, including underrepresented ancestries, rare or imbalanced disease cohorts, and restricted-access biobanks. Conditional synthetic genome generation would support privacy-aware benchmarking, simulation, and data augmentation, but requires learning an ultra-high-dimensional discrete distribution over hundreds of thousands to tens of millions of correlated single-nucleotide polymorphisms (SNPs), or sites of DNA variation. We introduce \textbf{HELICS}: \textbf{H}igh-fidelity g\textbf{E}neration via \textbf{L}atent \textbf{I}nterpolant \textbf{C}onditional \textbf{S}ynthesis, a phasing-free latent flow matching framework for genome-wide SNP-level generation from biobank-scale genotype data. HELICS tokenizes chromosomes with convolutional autoencoders, reducing \(>\)\(600{\rm K}\) SNPs to \(4{,}096\) continuous latent tokens, and trains a transformer parameterized conditional flow matching model over the concatenated multi-chromosome latent representation. Trained on over \(417\rm K\) genomes in the UK Biobank, HELICS generates genome-wide common-variant synthetic cohorts without haplotype phasing and achieves stronger fidelity to held-out real genotypes than existing haplotype- and genotype-based simulators, including high allele-frequency agreement in external evaluation (\(R^2=0.992\)) and superior preservation of local and long-range SNP correlation structure. Conditioning on ancestry and genetic liability summaries enables targeted generation across genetic risk profiles; in perturbation analyses, by increasing the conditioned disease trait liability, HELICS induces SNP-level dosage changes in generated genomes that are highly correlated with published association weights. HELICS provides a scalable generative modeling framework for SNP-level genome data and a path toward conditional synthetic cohorts for data-limited genetic studies.
\end{abstract}

\section{Introduction}

Modern genetic studies increasingly depend on large cohorts of individual-level genotype data, motivating major investments in biobanks for genome-wide association studies (GWAS), polygenic risk modeling, and therapeutic target discovery. However, collecting genotype data at the scale needed to answer crucial scientific questions remains slow and costly. Existing cohorts are highly imbalanced across ancestries, biasing discovery toward well-represented populations; many diseases are rare and have small sample sizes; and individual-level data are sensitive, identifiable, and access-controlled. These common data-scarce settings motivate synthetic genome generation not only as a privacy mechanism, but importantly, as a way to construct realistic, controllable genotypes to support method development, fairer and more powerful analyses of underrepresented ancestries, and even discovery of genetic mechanisms relevant to disease and therapeutic development.

Rather than modeling the four-letter DNA sequence directly, population-scale genetic analyses typically focus on specific sites of genomic variation known as single-nucleotide polymorphisms (SNPs). SNPs are the primary coordinates used to link genetic differences across individuals to disease risk, complex traits, and population history \cite{wellcome2007genome, visscher201710, international2005haplotype, patterson2006population}. Many genetic analyses focus on common variants (observed in >1\% of the population) as they are well-powered for association testing and capture substantial polygenic variation. SNPs are represented as dosage vectors \(\boldsymbol G\in\{0,1,2\}^{L}\), where each coordinate counts the number of copies of a specified allele carried at each variant position. This representation is compact relative to the full genome, but \(L\) still ranges from $10^5$ to $10^7$ correlated variants, and modern cohorts now include over $10^6$ participants, including UK Biobank (\(>500{\rm K}\)) \cite{bycroft2018uk}, FinnGen (\(>500{\rm K}\)) \cite{kurki2023finngen}, All of Us (\(>849{\rm K}\)) \cite{all2024genomic}, and the Million Veteran Program (\(>10^6\)) \cite{gaziano2016million}. Synthetic genome generation thus requires modeling an ultra-high-dimensional discrete distribution in which disease-relevant signal is often highly sparse \cite{yang2010common, boyle2017expanded}.

Recent progress in text, images, and proteins demonstrates that generative models for discrete data can scale when they exploit domain structure and learn compact representations of high-dimensional data \cite{brown2020language,ramesh2021zero,madani2023large}. SNP-level genomes present distinct challenges as they contain up to tens of millions of correlated discrete variants, and disease-relevant signal is distributed across many distant weak-effect loci. A practical generator must preserve genome-wide genetic data fidelity while capturing disease-relevant liability signals that can be used for conditional control in downstream analyses.

Existing generative formulations are not fully aligned with this target. 
Base-level genomic foundation models \cite{nguyen2023hyenadna, dalla2025nucleotide, zhou2023dnabert} are typically trained on reference genome sequences and are thus focused on regulatory or functional prediction tasks, not modeling human genetic variation. Classical haplotype simulators and reference-panel methods reproduce local genetic structure but typically rely on fixed reference cohorts or phased haplotypes, where a haplotype refers to the allele sequence carried on one inherited chromosomal copy. Phasing infers the two inherited haplotype sequences from unphased genotype dosages and remains computationally intensive at biobank scale, often requiring approximate iterative algorithms \cite{hofmeister2023accurate,browning2021fast}. Moreover, many methods target only small regions or GWAS-filtered variant sets, limiting genome-wide utility.

Here, we introduce \textbf{HELICS}: \textbf{H}igh-fidelity g\textbf{E}neration via \textbf{L}atent \textbf{I}nterpolant \textbf{C}onditional \textbf{S}ynthesis, a scalable latent flow matching framework for genome-wide SNP-level generation. HELICS directly models unphased biobank-scale genetic data, compresses SNP sequences into continuous latent tokens, and learns a genome-wide latent flow model that supports conditional sampling. We rigorously evaluate HELICS against existing genotype-generation approaches using downstream genetics benchmarks that assess fidelity to real cohorts and diversity of generated samples. Finally, we show that HELICS supports conditional generation, moving synthetic genome generation toward controllable simulation of biologically meaningful variation.

Our main contributions are:

\begin{enumerate}

    \item \textbf{Scalable phasing-free genome generation.} We introduce a genome-wide framework that tokenizes SNP dosages and models multi-chromosome latent sequence with a transformer-based flow matching model trained directly on unphased biobank data.

    \item \textbf{High-fidelity synthetic genomes for downstream genetic analysis.} HELICS generates synthetic cohorts that preserve key structure for downstream genetic analysis, including allele-frequency (AF) spectra, linkage disequilibrium (LD; local SNP correlation), ancestry geometry, and generated diversity.

    \item \textbf{Controllable generation of genetic-liability profiles.} HELICS conditions on ancestry and genetic-liability summaries to generate genomes with targeted polygenic profiles while retaining genome-wide realism. Perturbation of the conditional genetic liability induces SNP-level dosage changes that align with external GWAS effects, providing an interpretable check on conditional control.

\end{enumerate}

\section{Related Work}

\subsection{Synthetic genotype generators}

Synthetic genotype generation has traditionally relied on haplotype simulators and reference-panel methods such as HAPGEN2 \cite{su2011hapgen2}, Sim1000G \cite{dimitromanolakis2019sim1000g}, and HAPNEST \cite{wharrie2023hapnest}. Rather than learning a flexible distribution over genotype dosage data, these methods rely on phased reference panels. HAPGEN2 and HAPNEST generate genomes by recombining phased haplotypes in the observed reference panel, whereas Sim1000G estimates regional statistical moments and samples thresholded multivariate Gaussian variables. Learned genotype generators, including GAN- \cite{yelmenGAN} and diffusion-based approaches \cite{lampis2026snpgen, kenneweg2025generating}, provide more flexible modeling but are restricted to small genomic regions, single chromosomes, or small trait-associated SNP subsets. These restrictions limit their practical utility for biobank-scale synthetic data generation and can exclude fine-scale genotype structure such as weak polygenic signals outside predefined associated loci. Appendix~\ref{app:related-work} provides a detailed discussion of the assumptions and tradeoffs underlying each class of method. 

Conditional generation has been widely implemented in biomedical imaging \cite{frid2018gan, pombo2023equitable, hung2023med, isola2017image} and ``digital-twin'' clinical modeling \cite{shmatko2025learning, urbut2024aladynoulli}, where generative models are used to simulate samples under specified attributes or counterfactual conditions. Yet this area remains underdeveloped for genome-wide genotype generation. Existing trait-conditioned genotype models typically require a predefined trait-relevant SNP set \cite{lampis2026snpgen}, which limits their applicability to highly polygenic traits, where genetic risk is distributed across many variants of individually small effect, and to discovery settings, where the conditionally relevant variants are not known in advance. A critical open problem is to learn disease-relevant structures from genome-wide variation, rather than restricting generation to variants preselected by external association tests. Existing genotype generators do not address that setting which we target in this work, namely whole-genome controllable generation, learned directly from unphased biobank-scale dosage data.

\subsection{Base-level genomic sequence models}

A separate line of work has advanced base-level DNA modeling through long-context architectures, bidirectional objectives, and learned tokenization for nucleotide strings \cite{duan2025janusdna, zhou2023dnabert, dalla2025nucleotide, nguyen2023hyenadna}. These models are mainly designed for regulatory, functional, and molecular prediction tasks over sequences in \(\{A,C,G,T\}\). For population and disease genetics, the relevant object is often different as individuals share most of the reference genome. Phenotypic or ancestry differences are instead commonly studied at sites of human genetic variation, especially SNPs, represented as allele dosages across individuals. Because genome-wide association studies, polygenic risk scoring, ancestry inference, and many biobank-scale analyses are formulated at the variant level, we focus on SNP-level generation rather than base-level DNA sequence generation.

\subsection{Latent generative modeling for high-dimensional data}

High-dimensional generative modeling often relies on learned latent representations. Latent diffusion and flow models encode observations into a lower-dimensional space and learn a latent space transport map which reduces computational cost \cite{rombach2022high, vahdat2021score}. Recent high-compression autoencoders further show that latent quality and structure can strongly affect downstream generative fidelity \cite{chen2024deep, chen2025dc}. This strategy is especially useful for genome-scale SNP data, where direct discrete modeling would require operating over hundreds of thousands to millions of categorical coordinates. Autoregressive and discrete-diffusion models provide alternatives for discrete data \cite{van2016pixel,hoogeboom2021argmax,austin2021structured}, but autoregressive models impose a unidirectional factorization that is not clearly justified for genotype data, and direct categorical denoising is difficult to scale to tens of millions of SNPs. We thus combine flexible genotype encoders with flow matching in a learned continuous latent space.

\section{Method}

\subsection{Problem setup}
\label{sec:setup}

We represent each individual genome as a collection of chromosome-indexed variant dosage sequences. For individual \(i\) and chromosome \(c\in\{1,\ldots,22\}\), HELICS takes as input \(\boldsymbol{G}_{c,i} \in \{0,1,2\}^{L_c}\),  where \(L_c\) is the number of retained variants on chromosome \(c\), and each entry denotes the dosage at a given SNP. Variants are ordered by genomic position; unlike other biological data modalities where features have no meaningful ordering (e.g. single-cell or proteomic data), genomic variants are not exchangeable since local dependence is induced by LD and shared haplotypic structure. 
HELICS directly models unphased genotype dosages across the genome, unlike many existing approaches, and avoids a computationally expensive preprocessing step at biobank scale. 

Leveraging the wealth of data available in modern large-scale biobanks, we train our generative models on UK Biobank (UKB) genotype data. Use of UKB data was approved under application 52008. After data preprocessing and quality control (see Appendix \ref{app:UKB}), the data comprised \(604{,}243\) autosomal SNPs and \(427{,}948\) individuals across multiple ancestries, of which \(10{,}000\) individuals were held out as a validation set and the rest were used for training the HELICS multi-ancestry model. We also train a single-ancestry version of HELICS for which the training data comprised \(251{,}465\) unrelated self-reported White British individuals. 

\subsection{Latent token representation}
\label{sec:latentTokenRep}

\begin{figure}
    \centering
    \includegraphics[width=0.85\linewidth]{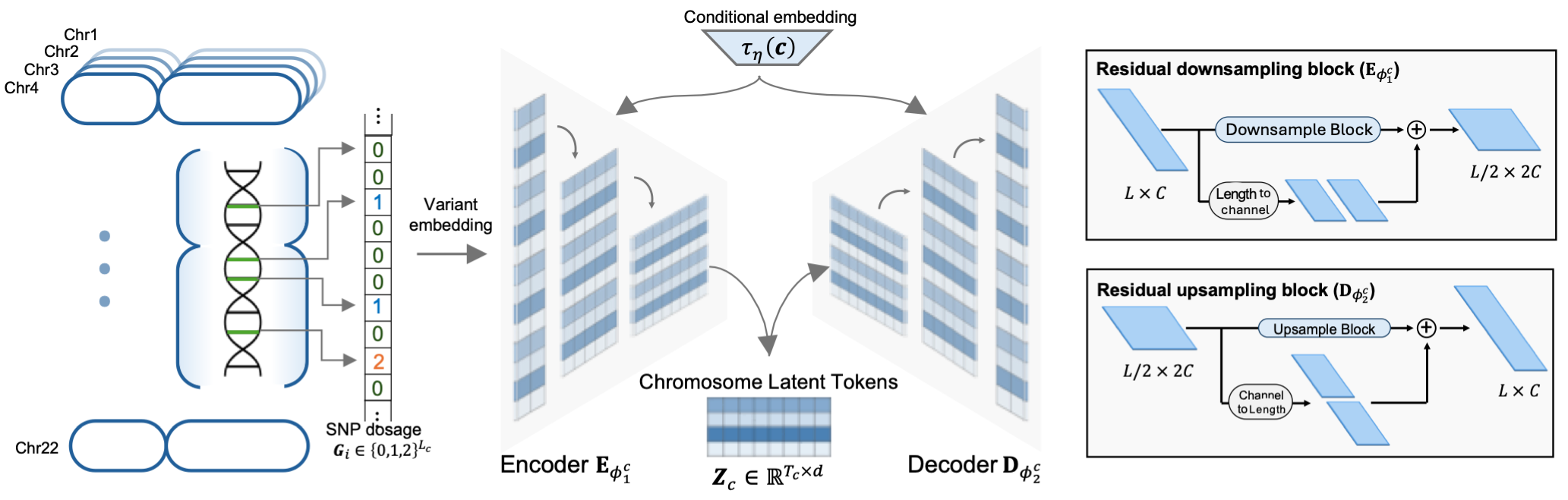}
    \caption{
    Chromosome-specific conditional autoencoder architecture.
    Unphased SNP dosages are encoded into chromosome-level latent tokens and decoded back to genotype logits (left) via residual downsampling and upsampling blocks (right).
    }
    \label{fig:AEarchitecture}
\end{figure}

HELICS adapts a transformer-based architecture for high-fidelity generative modeling. However, directly modeling SNP sequences with a transformer is challenging as the data are very high-dimensional (\(L>600K\) SNPs modeled here). 
To train generative models in this ultra-high-dimensional setting while capturing potentially non-linear and long-range interactions, we first train chromosome-specific autoencoders to embed genotype dosage data into a smaller set of continuous latent tokens of fixed dimension.
For each chromosome \(c\), SNP dosages are one-hot encoded and mapped by a chromosome-specific encoder \(\mathbf{E}_{\phi_1^c}: \{0,1\}^{L_c\times 3} \rightarrow \mathbb R^{T_c \times d}\) to continuous latent tokens \(\boldsymbol{Z}_{c,i} := \mathbf{E}_{\phi_1^c}(\boldsymbol{G}_{c,i})\), where \(T_c \ll L_c\) varies with chromosome length and \(d\) is shared across chromosomes. In our genome-wide model, \(\sum_{c=1}^{22}T_c = 4{,}096\) and \(d=256\). Recognize that this representation shortens each chromosome from \(L_c\) SNP positions to \(T_c\) tokens, which reduces downstream transformer attention from raw-position scaling in \(L_c^2\) to latent-token scaling in \(T_c^2\).

Each chromosome is zero-padded to its working length, mapped to a power-of-two resolution by a strided convolution, and encoded by residual downsampling blocks (Figure~\ref{fig:AEarchitecture}a); padding is removed after decoding, restoring the original chromosome length $L_c$. Each block sums a learned strided convolution branch with a non-parametric length-to-channel shortcut, analogous to pixel-unshuffle \cite{shi2016real}; the decoder mirrors this with residual upsampling, which is analogous to pixel-shuffle. Local LD structure motivates this convolutional design and provides an efficient tokenization strategy. Chromosome-specific resolutions, depths, token counts \(T_c\), and channel schedules are chosen by a precomputed architecture search under the genome-wide token budget; details are in Appendix \ref{app:AEarchitecture-details}.

Chromosome-specific decoders map latent tokens back to genotype logits, \(
\mathbf{D}_{\phi_2^c}:\mathbb R^{T_c\times d}\rightarrow \mathbb R^{L_c\times 3},\)
where the three logits correspond to genotype dosage classes \(0,1,2\). Let
\(
p_{c,ijg}
=
\mathrm{softmax}(\mathbf{D}_{\phi_2^c}(\boldsymbol{Z}_{c,i})_{j,:})_g
\)
for \(g\in\{0,1,2\}\) denote the decoded genotype probabilities for individual \(i\) and SNP \(j\) on chromosome \(c\), and let
\(
\widehat G_{c,ij}=\sum_{g=0}^{2} g\,p_{c,ijg}
\)
be the decoded expected dosage. During generation, hard genotype calls are obtained by
\(
\widetilde G_{c,ij}=\arg\max_{g\in\{0,1,2\}}p_{c,ijg}.
\)
For a minibatch \(\mathcal B\) and chromosome \(c\), the autoencoder is trained with
\[
\begin{aligned}
\mathcal L_{\mathrm{AE}}^{(c)}
=&
-\frac{1}{|\mathcal B|L_c}
\sum_{i\in\mathcal B}
\sum_{j=1}^{L_c}
\log p_{c,ij,G_{c,ij}}
+
\lambda_{\mathrm{MAF}}
\frac{1}{L_c}
\sum_{j=1}^{L_c}
\left|
\frac{1}{2|\mathcal B|}
\sum_{i\in\mathcal B}
\widehat G_{c,ij}
-
\frac{1}{2|\mathcal B|}
\sum_{i\in\mathcal B}
G_{c,ij}
\right|
\\
&+
\lambda_{\mathrm{LD}}
\frac{1}{|\mathcal W_c|}
\sum_{W\in\mathcal W_c}
\left\|
\mathrm{Corr}\!\left(\widehat{\boldsymbol{G}}_{c,\mathcal B,W}\right)
-
\mathrm{Corr}\!\left(\boldsymbol{G}_{c,\mathcal B,W}\right)
\right\|_F
+
\lambda_{\rm reg}\mathcal L_{\mathrm{reg}}^{(c)} 
+ 
\lambda_{\mathrm{cond}}\mathcal L_{\mathrm{cond}}^{(c)} 
\end{aligned}
\]
Here, the first term is SNP-wise cross-entropy, the second term penalizes the mean absolute difference between decoded and empirical minor allele frequencies (MAF), computed from expected dosages and tuned by hyperparameter $\lambda_{\rm MAF}$, and the third term penalizes local LD distortion, tuned by hyperparameter $\lambda_{\rm LD}$, where \(\mathcal W_c\) denotes the set of non-overlapping genomic windows on chromosome \(c\) used in the minibatch LD penalty, \(\boldsymbol{G}_{c,\mathcal B,W}\in\mathbb R^{|\mathcal B|\times |W|}\) and \(\widehat{\boldsymbol{G}}_{c,\mathcal B,W}\in\mathbb R^{|\mathcal B|\times |W|}\) are the empirical and decoded dosage matrices restricted to window \(W\), and \(\mathrm{Corr}(\cdot)\) is the sample correlation matrix across SNPs within that window. Recognize that these constraints penalize cohort-level distortions beyond reconstruction accuracy; for example, predicting the major genotype almost everywhere can yield high reconstruction accuracy while disregarding low-frequency alternate alleles. Crucially, allele frequency differences underlie ancestry-related population structure \cite{patterson2006population, price2006principal}, and LD directly affects variant tagging and polygenic-score behavior \cite{wray2005allele, vilhjalmsson2015modeling}. 

The condition-prediction term \(\mathcal L_{\mathrm{cond}}^{(c)}\) mean-pools latent tokens and predicts the conditioning vector with a two-layer MLP, thereby encouraging chromosome latents to retain ancestry- and liability-relevant information; see Appendix \ref{sec:app-AEobjective} for more details. Finally, the regularization term \(\mathcal L_{\mathrm{reg}}^{(c)}\) contains latent-stabilization penalties used to make the encoded representation more suitable for downstream flow matching; \(\lambda_{\rm reg}>0\) only during the final \(25\%\) of autoencoder training. We ablate these terms in Appendix \ref{sec:app-ablateLreg}.

\subsection{Latent-space flow matching}
\label{sec:latent-flow}

\begin{figure}
    \centering
    \includegraphics[width=\linewidth]{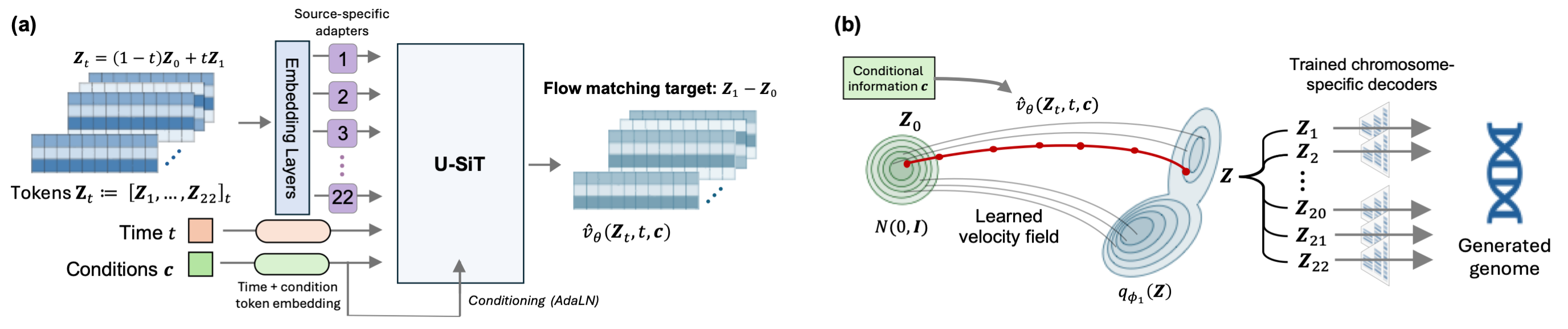}
    \caption{Latent flow matching for conditional genome generation. (a) A conditional USiT model architecture is used to learn a velocity field over concatenated chromosome latent tokens. (b) Synthetic genomes are generated by integrating over the learned velocity field, splitting generated tokens by chromosome and decoding with trained decoders.}
    \label{fig:flow}
\end{figure}

The autoencoder is frozen after the first stage of training, and a latent space flow matching model is trained. Let \(\boldsymbol Z_i=\{\boldsymbol Z_{1,i},\ldots,\boldsymbol Z_{22,i}\}\in\mathbb R^{T\times d}\) denote the concatenated chromosome tokens for individual \(i\), where \(T=\sum_{c=1}^{22} T_c\), and let \(q_{\phi_1}(\boldsymbol Z)\) be the empirical distribution of autoencoder latents induced by the set of trained encoders. We learn a transport map from a simple latent prior to \(q_{\phi_1}\) using flow matching \cite{lipman2022flow}. 
Specifically, we sample \(\boldsymbol Z_0\sim\mathcal N(0,\boldsymbol I)\), \(\boldsymbol Z_1\sim q_{\phi_1}(\boldsymbol Z)\), and \(t\sim\mathrm{Unif}(0,1)\), define \(\boldsymbol Z_t=(1-t)\boldsymbol Z_0+t\boldsymbol Z_1\), and train \(v_\theta(\boldsymbol Z_t,t)\) to predict the constant interpolation velocity \(v^\star(\boldsymbol Z_t,t)=\boldsymbol Z_1-\boldsymbol Z_0\), under the following loss:
\[
\mathcal L_{\mathrm{FM}}
=
\mathbb E_{t,\boldsymbol Z_0,\boldsymbol Z_1}
\left[
\left\|v_\theta(\boldsymbol Z_t,t)-v^\star(\boldsymbol Z_t,t)\right\|_2^2
\right],
\]
where \(v_\theta(\boldsymbol{Z}_t,t)\) is replaced with \(v_\theta(\boldsymbol Z_t,t,\boldsymbol c)\) when training conditional models. We parameterize \(v_\theta\) with USiT, an SiT-style latent transformer adapted from DC-AE generative modeling \cite{chen2024deep}, combining UViT \cite{bao2023all} with the SiT sampler \cite{ma2024sit}. USiT applies full self-attention over the concatenated multi-chromosome latent tokens with learned positional, chromosome, time, and condition embeddings, plus lightweight source-aware token adapters (Figure~\ref{fig:flow}a). Genome-wide experiments use a U-shaped variant with \(H\) encoder-to-decoder skip connections to preserve token-level information during global processing (Figure~\ref{fig:usit_block}); architectural details are in Appendix~\ref{app:usit_details}.

At generation time, we sample \(\boldsymbol Z_0\sim\mathcal N(0,\boldsymbol I)\) and integrate the learned vector field with DPM-Solver++ \cite{lu2025dpm} to obtain a generated latent $\widehat{\boldsymbol Z}$. Chromosome-specific decoders then partition and map \(\widehat{\boldsymbol Z}\) to make genotype calls (Figure~\ref{fig:flow}b). For multi-ancestry HELICS, we apply two additional post-training decoder refinement procedures; decoder-side refinement is a standard procedure in compressed latent models \cite{chen2025deep, yu2024image, chen2025masked}. In the latent space, generated tokens undergo spectral projection onto token-specific subspaces explaining \(99\%\) of training latent variance; this removes generated variation along low-variance token directions that are poorly supported by the training latent distribution. In the genotype space, genotype calls undergo a fixed low-rank PC alignment to the training genotypes. Importantly, spectral statistics are estimated from training latents, and PC alignment is fitted once using training genotypes and a fixed generated refinement sample. These refinement procedures are frozen and applied unchanged to every subsequent generation using no held-out evaluation data. These refinements form an essential final stage of the HELICS model-fitting procedure; the unrefined generator represents an intermediate state rather than an intended model endpoint. Reported multi-ancestry fidelity reflects the complete refined generator. Details and ablations are provided in Appendix~\ref{app:post_generation_refinement}.

\subsection{Controllable trait-guided generation}
\label{sec:trait-guided-generation}

We condition generation on target profiles \(\boldsymbol{c}_i\) comprising ancestry coordinates and discretized genetic liabilities to a set of preselected traits. In training, we compute polygenic risk scores (PRS) for each trait \(k\) and individual \(i\),
\(\mathrm{PRS}_{ik}=\sum_{j\in\mathcal S_k}\beta_{jk}G_{ij},\)
where \(\mathcal S_k\) denotes matched score variants, \(\beta_{jk}\) is the published effect size for variant \(j\), and \(G_{ij}\) is the effect-allele dosage. Trait selection, PRS sources, and preprocessing are detailed in Appendix~\ref{app:conditions}. We assign each PRS to a decile bin and use decile labels and ancestry coordinates given by the top 20 genetic principal components, as the conditioning profile. At generation time, specifying a target bin controls the desired genetic liability stratum. 

Conditioning is applied in both stages of the model. The autoencoder injects a learned embedding \(\tau_\eta(\boldsymbol{c}_i)\) via FiLM-style modulation \cite{perez2018film}, and the latent flow model learns \(v_\theta(\boldsymbol{Z}_t,t,\boldsymbol{c}_i)\) using DiT-style adaLN-Zero modulation and a concatenated condition token \cite{peebles2023scalable}. Main results use standard conditional sampling, equivalent to classifier-free guidance with scale \(1.0\) \cite{ho2022classifier}; guided sampling is evaluated in Appendix~\ref{app:cfg}. In Section~\ref{sec:trait_guided_results}, we assess control by conditioning on held-out validation profiles \(\boldsymbol{c}_i\) and comparing generated genotypes to their observed counterparts.

\section{Experiments and results}

\subsection{Evaluation setup}
\label{sec:evals}

We evaluate HELICS against the UKB \cite{bycroft2018uk}, the training source for HELICS; 1000 Genomes (1KG) \cite{auton2015global}, the reference panel used by several haplotype-based baselines; and NHGRI Centers for Common Disease Genomics (CCDG) \cite{regier2018functional}, an external population-scale cohort (\(N\approx 140\mathrm{K}\)). We consider EUR-ancestry benchmarks here. HELICS generates genome-wide data, but since several prior methods are limited to the regional panel of \cite{yelmenGAN}, benchmarks use harmonized common variants on chromosome 15 across UKB/1KG/CCDG after filtering to MAF \(>1\%\), yielding \(18{,}784\) SNPs; each method is evaluated on its available overlap with this harmonized SNP set.

We compare HELICS and HELICS-EUR (a single-ancestry model trained only on EUR UKB samples) to haplotype simulators HAPNEST \cite{wharrie2023hapnest}, HAPGEN2 \cite{su2011hapgen2}, and Sim1000G \cite{dimitromanolakis2019sim1000g}, and released learned generators based on GANs, restricted Boltzmann machines (RBMs) \cite{yelmenGAN}, and hierarchical latent tree (HCLT) models \cite{dang2023HCLTs}. Many existing generators are designed for restricted genomic panels or smaller reference cohorts and cannot readily leverage a full UKB training set. We hence evaluate these methods under their stated settings and emphasize that regional genotype generators constitute much of the existing literature. Scaling to both genome-wide variant panels and biobank-sized training cohorts is a central contribution of HELICS. Training-size and reference-panel sensitivity analyses and additional benchmarking details for all baseline methods are provided in Appendix~\ref{app:benchmarks}.

Metrics quantify genetic data fidelity, privacy proxies, and generated diversity. Allele-frequency fidelity compares generated and reference MAFs, namely the frequencies of the less common alleles, using \(R^2\) and MAE. LD fidelity compares local SNP correlation structure using the length-normalized Frobenius norm of the difference between generated and reference LD matrices (smaller values indicate closer LD). Population structure is assessed by projection into reference genotype-PC space. Membership signal measures whether nearest-neighbor (NN) distance from real individuals to the generated cohort separates UKB training from held-out validation samples, reported as \(|\mathrm{AUC}-0.5|\), with values near zero indicating little train-validation separability; this is a proxy for privacy. Generated NN IBS similarity measures within-synthetic nearest-neighbor identity-by-state (IBS) similarity, where lower values suggest more diversity and reduced mode collapse. More detailed definitions of these metrics and additional benchmarks are included in Appendix~\ref{app:benchmarks}.

\subsection{HELICS generates high-fidelity biobank-scale genomes}
\label{sec:genBench}

\begin{table}[t]
\centering
\caption{Chromosome 15 benchmarks against external CCDG cohort, with UKB and 1KG as reference baselines. GW: whether a method is designed for genome-wide generation. Each method uses its available overlap with the harmonized common chr15 reference panel. Low-MAF \(R^2\): \(R^2\) for \(0.01<\mathrm{MAF}<0.05\); LD diff.: Frobenius norm of the LD difference; Membership: \(|\mathrm{AUC}-0.5|\) for train-validation distinguishability by distance from synthetic set; Gen. NN IBS: generated nearest-neighbor IBS similarity.}
\scriptsize
\setlength{\tabcolsep}{3pt}
\renewcommand{\arraystretch}{0.92}
\begin{tabular}{l c r c c c c c c}
\toprule
\textbf{Approach}
& \textbf{GW}
& \textbf{No. SNPs}
& \textbf{MAF \(R^2\) \(\uparrow\)}
& \textbf{Low MAF \(R^2\) \(\uparrow\)}
& \textbf{MAF MAE \(\downarrow\)}
& \textbf{LD diff. \(\downarrow\)}
& \textbf{Membership \(\downarrow\)}
& \textbf{Gen. NN IBS \(\downarrow\)} \\
\midrule
HAPNEST \cite{wharrie2023hapnest}      & \checkmark & 10{,}509 & 0.942 & 0.207 & 0.025  & 0.101 & 0.030 & 0.914 \\
Sim1000G \cite{dimitromanolakis2019sim1000g} & \ding{55} & 9{,}899 & 0.988 & 0.637 & 0.011 & 0.122 & 0.011 & 0.839 \\
HAPGEN2 \cite{su2011hapgen2}      & \ding{55}  & 17{,}117 & 0.976 & 0.444 & 0.016  & 0.069 & 0.005 & 0.879 \\
GAN-based \cite{yelmenGAN}    & \ding{55}  &  6{,}513 & 0.947 & 0.129 & 0.023  & 0.129 & 0.060 & 0.923 \\
RBM-based \cite{yelmenGAN}   & \ding{55}  &  7{,}531 & 0.829 & 0.191 & 0.032  & 0.110 & 0.001 & 0.927 \\
HCLT-based  \cite{dang2023HCLTs}       & \ding{55}  &  6{,}827 & 0.846 & 0.140 & 0.036  & 0.152 & 0.029 & 0.906 \\
\textbf{HELICS} & \checkmark & 18{,}784 & 0.992 & 0.362 & 0.019 & 0.115 & 0.009 & 0.914 \\
\textbf{HELICS-EUR} & \checkmark & 18{,}784 & 0.962 & 0.493 & 0.020 & 0.067 & 0.004 & 0.847 \\
\midrule
UKB ref.     & -- & 18{,}784 & 0.998 & 0.950 & 0.004 & 0.014 & 0.004 & 0.839 \\
1000G ref.   & -- & 18{,}784 & 0.989 & 0.762 & 0.010 & 0.036 & 0.012 & 0.933 \\
\bottomrule
\end{tabular}
\label{tab:main_benchmark}
\end{table}

Table~\ref{tab:main_benchmark} reports chromosome 15 results against the external CCDG EUR reference. HELICS shows strong overall performance. The multi-ancestry model achieves the strongest overall MAF agreement, while HELICS-EUR achieves the lowest LD discrepancy and strong low-frequency MAF agreement, as expected under ancestry-matched training. Regional reference-based simulators such as Sim1000G and HAPGEN2 are included for completeness and remain competitive for low-frequency variants, but their unconditional regional sampling is substantially narrower in scope than HELICS's joint genome-wide, trait-conditioned generation from biobank-scale data. Local learned generators have fewer overlapping SNPs and are generally outperformed here. HELICS combines strong genetic fidelity with genome-wide generation from unphased dosages and trait conditioning.

Membership and diversity metrics show no evidence of substantial leakage or copying. HELICS models have membership deviations near reference baselines and comparable to the best-performing baselines, as well as NN IBS similarity close to the UKB control for HELICS-EUR. As an additional privacy-motivated check, we used KING \cite{manichaikul2010robust}, a genetic relatedness estimator, to test whether generated samples exhibit kinship with training individuals in the UKB. HELICS produced no first- or second-degree synthetic-to-training sample relationships, and its third-degree pair rate is below the matched validation-to-training rate (Table~\ref{tab:king_relatedness}), indicating no close kinship-level copying. See Appendix \ref{app:benchmark-metric-definitions} for more details.

\begin{figure}
    \centering
    \includegraphics[width=\linewidth]{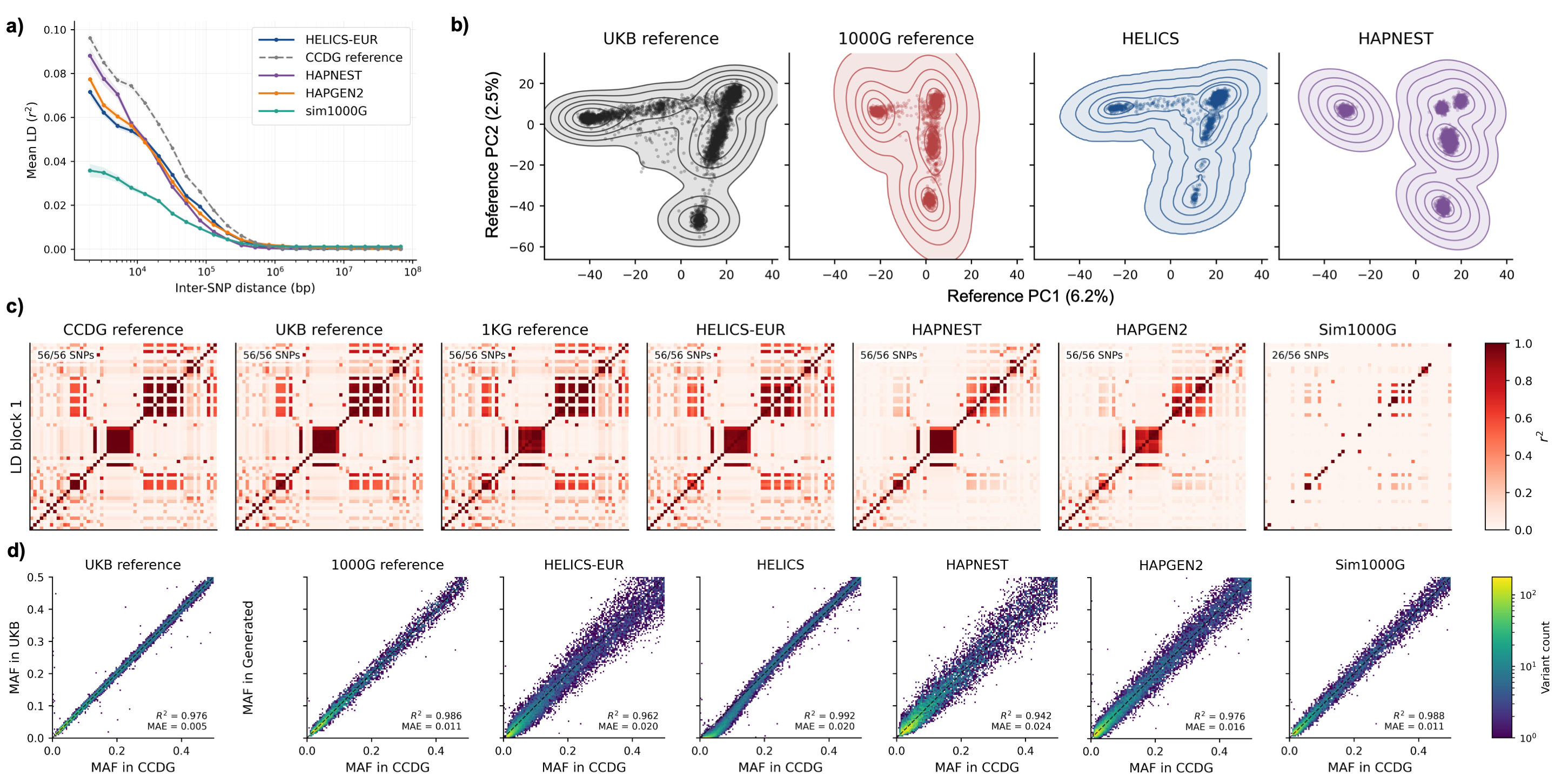}
    \caption{Synthetic genome fidelity on chromosome 15. Panels show (a) LD decay, (b) genotype-PC projections, (c) local LD heatmaps (chr15:24{,}321{,}414--24{,}848{,}545), and (d) MAF agreement. Restricted regional baselines are omitted from these chromosome-scale evaluations.}
    \label{fig:benchmarksCompare}
\end{figure}

Figure~\ref{fig:benchmarksCompare} qualitatively supports the chromosome 15 benchmark. We omit methods trained only on small regional panels because their limited variant overlap does not support chromosome-scale LD decay, PCA, or LD block comparisons. HELICS closely follows reference LD decay, including at longer distances (\(>10^4\) bp), while Sim1000G shows weaker short-range LD. In UKB-fit genotype-PC space, HELICS better preserves the continuous EUR reference geometry, whereas HAPNEST forms sharper clusters. Local LD heatmaps show that HELICS-EUR recovers block structure comparable to the reference panels, and MAF scatter plots show close agreement. While HELICS-EUR is more accurate at lower-frequency variants, multi-ancestry HELICS closely tracks reference MAFs above \(0.1\); the low-frequency discrepancy is attributable to discretization in the post-training refinement procedure, which is discussed in Appendix \ref{app:post_generation_refinement}. Overall, HELICS preserves both marginal and dependence structure while scaling beyond regional panels without phased-haplotype preprocessing.

\begin{table}[h]
\centering
\caption{Metrics and abbreviations follow Table~\ref{tab:main_benchmark}. Values are means \(\pm\) SD across 22 chromosomes or, for Sim1000G, independently simulated regions for each autosome.}
\label{tab:autosome_benchmark}
\scriptsize
\resizebox{\linewidth}{!}{%
\begin{tabular}{lcccccccc}
\toprule
\textbf{Approach} 
&\textbf{Scope}
& \textbf{No. SNPs}
& \textbf{MAF \(R^2\) \(\uparrow\)}
& \textbf{Low MAF \(R^2\) \(\uparrow\)}
& \textbf{MAF MAE \(\downarrow\)}
& \textbf{LD diff. \(\downarrow\)}
& \textbf{Membership \(\downarrow\)}
& \textbf{Gen. NN IBS \(\downarrow\)} \\
\midrule
HAPNEST \cite{wharrie2023hapnest} & Genome-wide & 349,564 & \(0.926\pm0.003\) & \(0.177\pm0.012\) & \(0.027\pm0.001\) & \(0.118\pm0.003\) & \(0.028\pm0.020\) & \(0.820\pm0.004\) \\
Sim1000G \cite{dimitromanolakis2019sim1000g} & Regional & \(4,973\times22\) & \(0.985\pm0.001\) & \(0.579\pm0.034\) & \(0.012\pm0.0004\) & \(0.127\pm0.026\) & \(0.020\pm0.017\) & \(0.843\pm0.006\) \\
\textbf{HELICS} & Genome-wide & 600,336 & \(0.980\pm0.009\) & \(0.489\pm0.199\) & \(0.023\pm0.006\) & \(0.109\pm0.013\) & \(0.013\pm0.006\) & \(0.893\pm0.027\) \\
\textbf{HELICS-EUR} & Genome-wide & 600,336 & \(0.976\pm0.009\) & \(0.502\pm0.069\) & \(0.023\pm0.003\) & \(0.107\pm0.025\) & \(0.030\pm0.010\) & \(0.866\pm0.013\) \\
\bottomrule
\end{tabular}}
\end{table}

Table~\ref{tab:autosome_benchmark} extends the comparison across all 22 autosomes, with chromosome-resolved results in Appendix Figure~\ref{fig:ancestry_fidelity_chromosomes}. We select HAPNEST and Sim1000G as representative reference-based simulators for this broader comparison; local learned generators were trained primarily on restricted SNP panels and were thus excluded from this all-autosome comparison. Note that while HELICS and HAPNEST generate genome-wide samples, Sim1000G is evaluated on independently simulated regions. Sim1000G's stronger regional MAF matching is consistent with its reference-based formulation, but it does not support joint genome-wide generation or trait conditioning. HELICS supports these capabilities and achieves lower mean LD discrepancy across autosomes. For each method, evaluated SNP count reflects overlap with the harmonized benchmark reference panel. 

\subsection{Latent space captures human genetic structure}
\label{sec:latent_structure}

We next characterize the learned latent space and find that its dominant geometry reflects genetic ancestry structure. Each individual is represented by a latent tensor \(\boldsymbol Z_i\in\mathbb R^{T\times d}\). For visualization, we standardize each token-coordinate across individuals, compute normalized Frobenius distances between latent tensors, and apply UMAP \cite{mcinnes2018umap} with the resulting distance matrix. Note that UMAP is used only as a qualitative visualization of this tokenized latent geometry.

\begin{figure}[b]
    \centering
    \includegraphics[width=\linewidth]{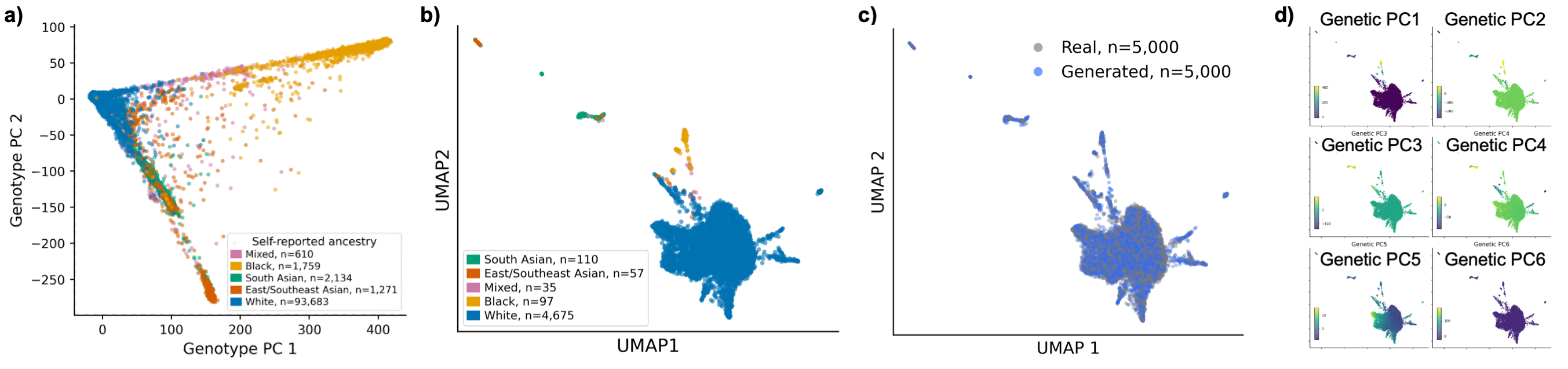}
    \caption{Genome-wide latents preserve human genetic structure. (a) UKB genotype PCs for \(100{,}000\) individuals colored by self-reported ancestry. (b) UMAP of held-out validation latents (\(N=5{,}000\)) using token-standardized Frobenius distances. (c) Real and generated latents overlap in the same embedding. (d) The embedding colored by genotype PCs shows preserved axes of genetic variation.}
    \label{fig:latentViz}
\end{figure}

Figure~\ref{fig:latentViz} shows that the learned latent space preserves major axes of human genetic variation. Held-out individuals organize by self-reported ancestry in the latent space, and genotype PCs vary continuously across this embedding. HELICS also shows strong coverage of the population variation on the UMAP coordinates, indicating that the model captures diverse ancestries in heterogeneous populations like the UKB. Chromosome-specific visualizations and coverage metrics in Appendix \ref{app:additional-res} show similar real/generated correspondence. We quantify latent alignment within self-reported ancestry groups using sliced \(W_2\) with 512 random projections and RBF MMD\(^2\) in the top 250 latent-PC space (Table~\ref{app:tab:latent_alignment}). Alignment is strongest for the majority White group and weaker for underrepresented groups, consistent with their smaller training sample sizes. Ancestry-stratified genotype evaluation likewise shows weaker fidelity in underrepresented groups (Table~\ref{tab:ancestry_fidelity}).

\subsection{Genetic liability condition perturbations reveal interpretable SNP-level changes}
\label{sec:trait_guided_results}

\begin{wrapfigure}{r}{0.35\linewidth}
    \vspace{-1.25em}
    \centering
    \includegraphics[width=\linewidth]{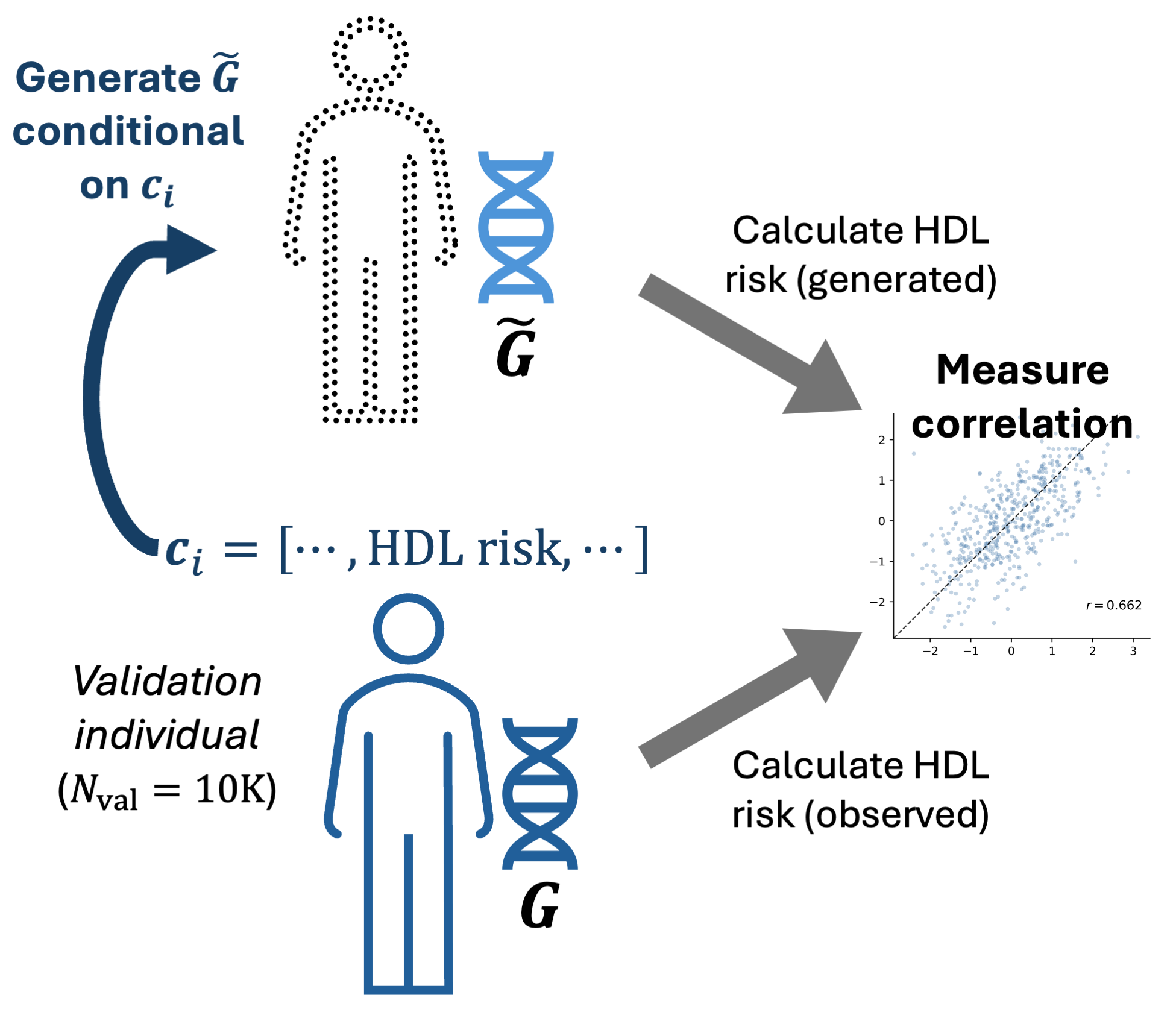}
    \caption{
    Conditional PRS control schematic. HELICS generates $\widetilde{\boldsymbol G}_i$ conditional on a validation profile $\boldsymbol c_i$. High correlation between generated and observed PRS indicates effective conditional control.
    }
    \label{fig:prs_hdl_schematic}
    \vspace{-1.0em}
\end{wrapfigure}
We next ask whether HELICS captures trait-relevant SNP-level genetic structure beyond ancestry and aggregate PRS control. To do so, we first assess conditional control using the procedure illustrated in Figure~\ref{fig:prs_hdl_schematic}. We focus on high-density lipoprotein (HDL) liability; HDL is a clinically relevant blood-lipid trait involved in cardiovascular risk, making it a useful test case for interpretable genetic-liability control. For each held-out validation individual \(i\), we use HELICS to generate \(\widetilde {\boldsymbol G}_i\) from that individual's conditioning profile \(\boldsymbol c_i\), including ancestry PCs and trait PRS deciles. We then compute HDL PRS from \(\widetilde {\boldsymbol G}_i\) and compare it to the HDL PRS of the corresponding observed genotype \(\boldsymbol G_i\). The generated and observed HDL PRSs are highly correlated (\(r=0.662\)), indicating effective conditional control as the conditional signal for HDL liability is preserved in the generated genomes. Further, rescoring with a separate HDL score not used for conditioning shows that positive generated-observed correlation persists after filtering variants in strong LD with the HDL conditioning-score variants (Appendix~\ref{app:external_hdl}).

\begin{figure}
    \centering
    \includegraphics[width=\linewidth]{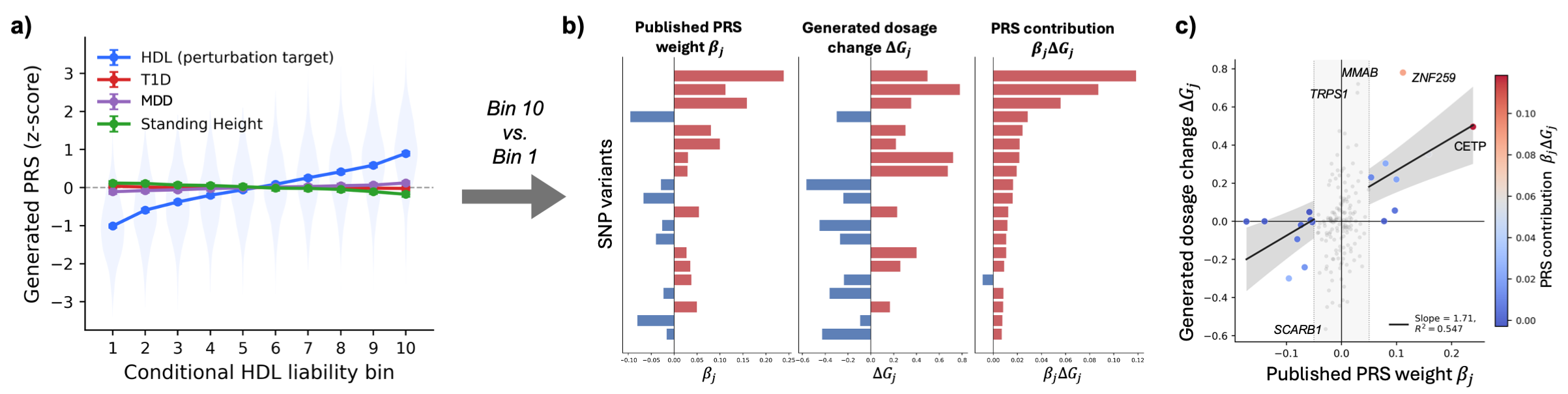}
    \caption{
    HDL conditioning produces association-consistent SNP-level responses. (a) Trait-guided perturbed generation controls HDL polygenic profiles with minimal off-target effects.
    (b) Among the largest HDL PRS-contributing shifts, most generated dosage changes \(\Delta G_j\) move in the direction predicted by the published HDL weight \(\beta_j\), yielding positive contributions \(\beta_j\Delta G_j\).
    (c) Across HDL score variants, bin-10 versus bin-1 dosage shifts are positively associated with published HDL weights, with large effects at known lipid loci. 
    }
    \label{fig:modelInterp}
\end{figure}

We next perturb HDL liability in the conditional information from low (bin 1) to high (bin 10) genetic liability, while holding ancestry and other trait conditions fixed; the goal is to isolate HELICS's response to the changing liability condition and then assess biological plausibility. Sweeping only the HDL PRS decile from 1 to 10 in the conditional information, we find that generated HDL PRS increases monotonically, while off-target PRS for Type 1 diabetes (T1D), major depressive disorder (MDD), and standing height remain comparatively stable (Figure~\ref{fig:modelInterp}a).

We explore further the mechanism by which HELICS modifies generated genomes to shift HDL liability from low to high. For the target trait HDL, let \(\widetilde {G}_j(10)\) and \(\widetilde {G}_j(1)\) denote generated dosages under bin 10 and bin 1 PRS conditioning, with all other conditions fixed. We define
\(
\Delta G_j =
\mathbb E[\widetilde G_j(10)]
-
\mathbb E[\widetilde G_j(1)],
\)
and compare this perturbation-induced dosage shift to the published PRS weight \(\beta_j\). Association-consistent conditional responses satisfy \(\beta_j\Delta G_j>0\) so that positive-effect alleles increase under high-bin conditioning and negative-effect alleles decrease. Note that HELICS is trained only on discretized PRS summaries and is never given variant-level weights \(\beta_j\).

We find that across the variants included in the HDL PRS, \(\Delta G_j\) is positively associated with \(\beta_j\) for high-effect variants (\(R^2=0.547\)), and the largest PRS-contributing shifts include known lipid-associated loci such as \textit{CETP}, \textit{MMAB}, \textit{TRPS1}, \textit{ZNF259}, and \textit{SCARB1}. Perturbing HDL conditioning induces SNP-level dosage shifts aligned with published HDL association effects, rather than producing uninterpretable aggregate PRS artifacts. This suggests that HELICS learns condition-dependent genotype variation at trait-relevant loci from genome-wide genotypes and score-derived labels, without receiving variant-level association weights.

\begin{wraptable}{r}{0.40\textwidth}
\vspace{-1.0em}
\centering
\caption{
Conditional-control summary across HDL, LDL, T1D, rheumatoid arthritis (RA), and apolipoprotein B (ApoB). 
For each trait, we report the number of variants included in the PRS, validation-profile PRS recovery (Gen--Obs. \(r\); see Figure~\ref{fig:prs_hdl_schematic}), low-to-high perturbation success (Bin 10 \(>\) 1), and contribution-weighted directional alignment among the top 20 SNP contributors to PRS shift.
}
\label{tab:conditional_control}
\scriptsize
\setlength{\tabcolsep}{3pt}
\renewcommand{\arraystretch}{0.95}
\begin{tabular}{lcccc}
\toprule
Trait & Variants & Gen-Obs. \(r\) & Bin 10 \(>\) 1 & Top-20 \\
\midrule
HDL  & 168     & 0.66 & 0.98 & 0.98  \\
LDL  & 606     & 0.75 & 1.00 & 0.94 \\
T1D  & 16{,}570 & 0.63 & 1.00 & 1.00 \\
RA   & 14{,}728 & 0.71 & 0.99 & 1.00  \\
ApoB & 43{,}001 & 0.71 & 0.95 & 0.80 \\
\bottomrule
\end{tabular}
\vspace{-1.0em}
\end{wraptable}
Table~\ref{tab:conditional_control} applies this perturbation analysis to additional traits (also see Figure~\ref{fig:modelInterp-othertraits}). The multi-ancestry HELICS model is used here; ancestry-stratified results are provided in Appendix Table~\ref{tab:conditional_control_ancestry}. Gen-Obs. \(r\) measures validation-profile PRS recovery, namely the correlation between generated and observed PRS across matched held-out individuals, as illustrated in Figure~\ref{fig:prs_hdl_schematic}. Bin 10 \(>\) 1 measures perturbation success, namely the fraction of profiles whose generated PRS is larger under decile-10 than decile-1 conditioning, as in Figure~\ref{fig:modelInterp}a. Top-20 measures SNP-level interpretability, specifically the contribution-weighted fraction of the 20 largest PRS-contributing dosage shifts satisfying \(\beta_j\Delta G_j>0\), as in Figure~\ref{fig:modelInterp}b. Even across highly polygenic traits like ApoB (\(43{,}001\) matched variants), HELICS controls genetic liability in response to conditional information, responds monotonically to low-to-high liability perturbations, and produces association-consistent dosage changes among the largest SNP-level responses.

\section{Conclusion}

We introduced HELICS, a scalable latent flow matching framework for SNP-level genome generation trained directly on unphased biobank data. By combining chromosome-specific encoding with a multi-chromosome latent flow, HELICS enables high-fidelity genome-wide synthesis at a scale beyond many existing genotype generators, preserving AF spectra, LD, ancestry structure, and sample diversity. Trait-guided experiments further show that perturbing genetic-liability profiles induces dosage shifts aligned with published association effects, despite the high dimensionality and polygenicity of the target traits.

Several directions remain to extend HELICS. Augmenting association studies of underrepresented cohorts or difficult-to-measure traits is an important next step, but requires robust statistical frameworks that preserve inferential validity and control inflation under synthetic data augmentation \cite{abdel2026harnessing}. HELICS is currently limited to common variants; future work should also extend HELICS to denser whole-genome panels including rare variants. Fidelity is weaker for low-frequency variants, and underrepresented ancestry groups show poorer latent alignment and PRS recovery. Transfer-learning strategies can be incorporated for underrepresented ancestries and imbalanced disease cohorts. 

Our perturbation analyses demonstrate that HELICS learns trait-associated patterns of genotype variation from genome-wide data and coarse liability summaries. This motivates further investigations into how HELICS represents shared genetic variation across traits and among distant loci. Further, conditioning on measured phenotypes and clinical covariates could support discovery of new genetic associations. These extensions can also enable sharing synthetic individual-level data for richer cross-cohort meta-analyses beyond marginal association summaries, provided sufficient fidelity, privacy, and inferential calibration can be achieved. Overall, conditional synthetic genome generation offers both a demanding use case for scalable generative modeling of structured discrete data and a promising tool for benchmarking, privacy-aware access, and controlled genetic simulation.

\section*{Acknowledgments and Disclosure of Funding}

This work was supported by grants from the National Institutes of Health R35-CA197449, R01-HL163560, U01-HG012064, U19-CA203654, P30-ES000002, T32-GM135117, and T32-CA009337. The authors are grateful to the participants and staff of the 1000 Genomes Project and the UK Biobank for their contributions. We acknowledge the participants, investigators, and sequencing centers of the Centers for Common Disease Genomics (CCDG) for generating and sharing the data used in this study. The CCDG program is supported by the National Human Genome Research Institute and the National Heart, Lung, and Blood Institute of the National Institutes of Health.

{
\small
\bibliography{references}
\bibliographystyle{unsrt}
}

\newpage
\appendix

\renewcommand{\thesection}{\Alph{section}}

\counterwithin{figure}{section}
\counterwithin{table}{section}
\counterwithin{equation}{section}

\renewcommand{\thefigure}{\thesection\arabic{figure}}
\renewcommand{\thetable}{\thesection\arabic{table}}
\renewcommand{\theequation}{\thesection\arabic{equation}}

\section*{Organization of the Appendix}

The appendix is organized as follows. We first describe data construction in Appendix~\ref{app:data-prep}, including UK Biobank preprocessing, external reference cohorts, ancestry assignment, and variant harmonization. In Appendix~\ref{app:architecture-details}, we describe the architectural and training details for HELICS, from chromosome-specific autoencoders to the latent flow matching model. Appendix~\ref{app:post_generation_refinement} describes the post-training refinement procedure used for multi-ancestry generation and ablates its components. Appendix~\ref{app:conditions} defines the conditioning variables used for controlled generation and examines classifier-free guidance. We then review existing synthetic genome generators in Appendix~\ref{app:related-work} and describe the benchmark construction, baseline generation procedures, and evaluation metrics in Appendix~\ref{app:benchmarks}. Finally, Appendix~\ref{app:additional-res} reports additional fidelity and relatedness results.

\section{Data preparation}
\label{app:data-prep} 
\subsection{UK Biobank data processing}
\label{app:UKB}

We used UK Biobank autosomal SNP genotype data derived from array-based genotyping and imputation. Starting from 488,377 individuals from the filtered data release, we retained variants on chromosomes 1-22 and applied sample and variant quality control using PLINK \cite{chang2015second}: individual missingness \texttt{---mind 0.02}, variant missingness \texttt{---geno 0.20}, minor allele frequency \texttt{---maf 0.01}, and Hardy-Weinberg equilibrium \texttt{---hwe 1e-12}. After quality control, the dataset comprised 427,948 individuals and 604,243 variants, with an overall genotyping rate of 0.9911.

We reserved 10,000 individuals for held-out testing and benchmarking. From the remaining samples, we used 251,465 unrelated British White individuals to train HELICS-EUR. An additional 42,669 British White individuals were retained for validation analyses, including hyperparameter tuning, phenotype-association evaluations, and exploratory analyses. 

We provide detailed methodological descriptions and have released implementation code at \texttt{https://github.com/ahmadazim/HELICS}. Due to the sensitivity of UKB genomic data release regulations, the training data and trained models are not publicly available. A public-data example is included; the demo downloads 1KG-EUR chr22, trains HELICS end to end, generates synthetic genotypes, and evaluates generated data fidelity.

\subsection{External reference cohorts: 1000 Genomes and CCDG}

We used two external multi-ancestry reference cohorts for benchmarking and ancestry harmonization: the 1000 Genomes Project (1KG) \cite{auton2015global} and the NHGRI Centers for Common Disease Genomics (CCDG) \cite{regier2018functional}. The 1000 Genomes Project Phase 3 release comprises 2,504 individuals from 26 populations and was carefully designed as a broad reference panel for human genetic variation across several continental ancestry groups \cite{auton2015global}. We include the 1KG cohort here since many of the baseline methods we benchmarked either use 1KG as their reference panel or require phased reference haplotypes. Since the 1KG Phase 3 haplotypes are publicly available in phased form, methods requiring phased input were run directly from the provided phased release without additional phasing.

We include the CCDG cohort as an external validation set to assess generalization beyond the training source or reference panel of all methods. CCDG is a large-scale NHGRI genome-sequencing effort designed to identify variants contributing to common disease phenotypes, with approximately \(140\rm{K}\) whole genomes sequenced \cite{regier2018functional}. Although CCDG samples were collected through disease-focused studies largely based in the United States, we did not rely on study labels or self-reported ancestry; 
genetic ancestry was inferred by projection into a 1KG-defined genetic-PC reference space. Briefly, we fit genotype PCs on 1KG reference samples with known superpopulation labels, projected each target sample into this PC space, and assigned ancestry by nearest-neighbor matching to the projected 1KG samples based on the top 20 PCs. This retained the 43,828 CCDG samples assigned closest to EUR ancestry for the EUR benchmark.

We used the 1KG cohort as the ancestry reference since it is a highly curated diverse reference panel with standardized population labels, as opposed to self-reported ancestry. Accordingly, 1KG was used to define the reference ancestry space for assigning EUR subsets, as described above, across generated samples, baseline methods, UKB, and CCDG.

\subsection{Variant harmonization}
\label{app:variant-harmonization}

Genotype data and baseline outputs were harmonized to a common chromosome and allele representation before benchmarking. We lifted all datasets to a common genome build (GRCh37) using \texttt{bcftools liftover} \cite{genovese2024bcftools}. After liftover, we retained autosomal biallelic common SNPs present across the reference datasets and matched variants by target-build position and unordered allele pair. The UKB data were used to define the canonical coordinate system and A1/A2 allele coding. For each reference or generated dataset, we applied A1/A2 recoding, allele swaps, and strand flips as needed to match the UKB convention; ambiguously matched variants were excluded. We then subset each benchmarked method to its overlap with this harmonized reference set. For chromosome 15, for example, this procedure produced a final benchmark reference panel of \(18{,}784\) harmonized variants.

\section{Architectural details}
\label{app:architecture-details}

\subsection{Chromosome-specific autoencoders}
\label{app:AEarchitecture-details}

HELICS trains one convolutional autoencoder per chromosome to tokenize the input genome-wide data. For chromosome \(c\), the encoder maps an unphased dosage sequence
\(
\boldsymbol G_c \in \{0,1,2\}^{L_c}
\)
to a sequence of continuous latent tokens
\(
\boldsymbol Z_c \in \mathbb R^{T_c \times d},
\)
where \(L_c\) is the number of retained SNPs on chromosome \(c\), \(T_c\) is the chromosome-specific latent token count, and \(d=256\) is the shared latent token dimension. The decoder then maps \(\boldsymbol Z_c\) back to per-SNP genotype logits over the three dosage classes \(\{0,1,2\}\). All chromosome-specific latents are concatenated before training the genome-wide latent flow model.

\subsubsection{Chromosome-specific token allocation and architecture search}

The total genome-wide latent token budget is fixed to
\(
T=\sum_{c=1}^{22}T_c=4096.
\)
Token counts are allocated across chromosomes in proportion to variant counts on each chromosome, while also constraining each \(T_c\) to be a power of two. Let \(w_c=L_c\) denote the allocation weight for chromosome \(c\), and define the continuous proportional target
\[
t_c = T \frac{w_c}{\sum_{c'=1}^{22} w_{c'}}.
\]
We choose token counts \(T_c\) by solving
\[
\min_{T_1,\ldots,T_{22}}
\sum_{c=1}^{22} |T_c-t_c|
\quad
\mathrm{s.t.}
\quad
T_c \in \{2^k:T_{\rm min}\le 2^k \le T_{\rm max}\},
\quad
\sum_{c=1}^{22}T_c=T,
\]
where $T_{\rm min}$ and $T_{\rm max}$ denote the minimum and maximum possible token allocation per chromosome, respectively. Here, we simply set \(T_{\rm min} = 64\) and \(T_{\rm max} = 1024\). The discrete optimization is solved exactly by dynamic programming over chromosomes and cumulative token count. Ties are broken deterministically by preferring the smaller token allocation and then the smaller previous cumulative sum. This gives a reproducible globally consistent allocation, avoiding independent rounding choices that generally fail to sum to the required total budget.

To choose a chromosome-specific sequence resolution that can be compressed exactly to the assigned token count \(T_c\), we first map the SNP sequence to a working length that can be reduced by one \(m_c\)-strided convolution followed by a fixed number of factor-two downsampling blocks. Specifically, in the first step of the encoder, we map the sequence to a working length of the form
\[
L_c^{\mathrm{work}} = m_c 2^{k_c},
\]
where \(m_c \in \{1,\ldots,m_{\max}\}\); here we take \(m_{\rm max} = 5\). 
The pair \((m_c,k_c)\) is selected so that \(L_c^{\mathrm{work}}\) is the smallest feasible length satisfying \(L_c^{\mathrm{work}}\ge L_c\) and permitting exact compression to \(T_c\). Specifically, an initial convolution with kernel size and stride \(m_c\) reduces the length from \(m_c2^{k_c}\) to \(2^{k_c}\), after which repeated factor-two downsampling reduces the sequence to \(T_c\). If \(L_c^{\mathrm{work}}>L_c\), the embedded sequence is zero-padded before encoding and truncated back to \(L_c\) after decoding. The number of residual downsampling stages is thus
\[
N_c^{\mathrm{down}} = k_c-\log_2 T_c.
\]
Each chromosome-specific architecture is determined by \(L_c\), \(T_c\), \(L_c^{\mathrm{work}}\), \(m_c\), and \(N_c^{\mathrm{down}}\) (Table~\ref{tab:ae_chrom_configs}).

\begin{table}[t]
\centering
\caption{
Chromosome-specific autoencoder configurations. \(L_c\) is the number of retained SNPs, \(L_c^{\mathrm{work}}\) is the padded working length, \(T_c\) is the allocated latent token count, and \(N_c^{\mathrm{down}}\) is the number of residual factor-two downsampling blocks.
}
\label{tab:ae_chrom_configs}
\small
\setlength{\tabcolsep}{4pt}
\renewcommand{\arraystretch}{0.95}
\begin{tabular}{lccccc}
\toprule
Chromosome & \(L_c\) & \(L_c^{\mathrm{work}}\) & \(m_c\) & \(T_c\) & \(N_c^{\mathrm{down}}\) \\ 
\midrule
1 & 47,555 & 49,152 & 3 & 256 & 6 \\ 
2 & 48,530 & 49,152 & 3 & 256 & 6 \\ 
3 & 40,846 & 40,960 & 5 & 256 & 5 \\ 
4 & 38,621 & 40,960 & 5 & 256 & 5 \\ 
5 & 36,643 & 40,960 & 5 & 256 & 5 \\ 
6 & 42,638 & 49,152 & 3 & 256 & 6 \\ 
7 & 33,158 & 40,960 & 5 & 256 & 5 \\ 
8 & 31,340 & 32,768 & 1 & 256 & 7 \\ 
9 & 26,650 & 32,768 & 1 & 128 & 8 \\ 
10 & 30,253 & 32,768 & 1 & 256 & 7 \\ 
11 & 29,723 & 32,768 & 1 & 256 & 7 \\ 
12 & 28,823 & 32,768 & 1 & 256 & 7 \\ 
13 & 20,930 & 24,576 & 3 & 128 & 6 \\ 
14 & 19,565 & 20,480 & 5 & 128 & 5 \\ 
15 & 19,011 & 20,480 & 5 & 128 & 5 \\ 
16 & 21,213 & 24,576 & 3 & 128 & 6 \\ 
17 & 19,643 & 20,480 & 5 & 128 & 5 \\ 
18 & 18,221 & 20,480 & 5 & 128 & 5 \\ 
19 & 16,333 & 16,384 & 1 & 128 & 7 \\ 
20 & 15,806 & 16,384 & 1 & 128 & 7 \\ 
21 & 9,080 & 10,240 & 5 & 64 & 5 \\ 
22 & 9,661 & 10,240 & 5 & 64 & 5 \\ 
\bottomrule
\end{tabular}
\end{table}

\subsubsection{Encoder and decoder blocks}

The input dosage vector is one-hot encoded giving an input tensor in \(\mathbb R^{B\times 3\times L_c}\). A \(1\times 1\) convolution maps this representation to an embedding dimension \(d_e\); in all experiments, \(d_e=8\). The sequence is then padded to \(L_c^{\mathrm{work}}\) as needed, and the first encoder operation reduces the working length from \(m_c2^{k_c}\) to \(2^{k_c}\) using a convolution with kernel size and stride \(m_c\). The encoder then applies \(N_c^{\mathrm{down}}\) residual downsampling blocks. As shown in Figure~\ref{fig:AEarchitecture}, given an activation \(h\in\mathbb R^{B\times C\times L}\), with batch size $B$, channel dimension $C$, and length $L$, each block maps
\[
(B,C,L) \mapsto (B,2C,L/2).
\]
The learned branch is a stride-two convolution, and the shortcut branch is a non-parametric one-dimensional pixel-unshuffle operation that rearranges adjacent length positions into channels. For an input activation \(h\in\mathbb R^{B\times C\times L}\),
\[
h_{\mathrm{out}}
=
f_{\mathrm{conv}}(h)
+
f_{\mathrm{unshuffle}}(h).
\]
where \(f_{\mathrm{conv}}\) is the learned convolution branch and \(f_{\mathrm{unshuffle}}\) is the non-parametric length-to-channel rearrangement shortcut. This shortcut preserves local information during downsampling without using an identity map, which is not shape-compatible when both length and channel dimension change.

The autoencoder is conditional. A condition vector \(\boldsymbol c_i\) is embedded with a two-layer MLP, and the resulting condition embedding modulates convolutional activations using FiLM-style affine transformations \cite{perez2018film}. For an activation \(h\), the modulation has the form
\[
\mathrm{FiLM}(h,\boldsymbol c)
=
h + \gamma(\boldsymbol c)\,\mathrm{GN}(h)+\beta(\boldsymbol c),
\]
where \(\mathrm{GN}\) is group normalization without affine parameters and \(\gamma(\boldsymbol c),\beta(\boldsymbol c)\in\mathbb R^{B\times C}\). The FiLM projection is initialized to zero, so conditional modulation is identity-preserving at initialization. Selected early encoder and decoder stages also use condition-dependent position gates, which produce a scalar gate per sequence position and multiply the activation pointwise. 

After the set of residual downsampling blocks, the sequence length is exactly \(T_c\). If the resulting channel count differs from the target latent dimension \(d\), a final channel-adjustment module maps the representation to \(d=256\) channels.

The decoder mirrors the encoder, inverting the channel-adjustment step if necessary and then applying \(N_c^{\mathrm{down}}\) residual upsampling blocks. Each upsampling block maps
\[
(B,C,L) \mapsto (B,C/2,2L).
\]
The learned branch is a transposed convolution, and the shortcut branch is the inverse one-dimensional pixel-shuffle rearrangement. Namely, for an activation \(h\),
\[
h_{\mathrm{out}}
=
f_{\mathrm{deconv}}(h)
+
f_{\mathrm{shuffle}}(h).
\]
where \(f_{\mathrm{deconv}}\) is the learned transposed-convolution branch and \(f_{\mathrm{shuffle}}\) is the non-parametric channel-to-length rearrangement shortcut. After the upsampling stack, a transposed convolution with kernel size and stride \(m_c\) maps the sequence back to \(L_c^{\mathrm{work}}\), and the output is truncated back to the original chromosome length \(L_c\) if padding was introduced before encoding. The final decoder layer is a \(1\times 1\) convolution from the embedding dimension to three genotype logits per SNP:
\(
\mathbf D_{\phi_2^c}(\boldsymbol Z_c)\in\mathbb R^{B\times L_c\times 3},
\)
where the three logits correspond to dosages \(g\in\{0,1,2\}\). We define
\(
p_{cijg}
=
\mathrm{softmax}\!\left(\mathbf D_{\phi_2^c}(\boldsymbol Z_{c,i})_{j,:}\right)_g
\)
and use the decoded expected dosage
\(
\widehat G_{cij}
=
\sum_{g=0}^{2} g\,p_{cijg}
\)
for differentiable reconstruction, allele-frequency, and LD-based training losses. For generated datasets and discrete benchmark outputs, hard genotype calls are obtained by
\[
\widetilde G_{cij}
=
\arg\max_{g\in\{0,1,2\}} p_{cijg}.
\]

\subsubsection{Autoencoder training objective}
\label{sec:app-AEobjective}

For chromosome \(c\), the autoencoder is trained to reconstruct genotype dosages while preserving chromosome-level allele-frequency structure, local LD, and condition-relevant information in the latent representation. Let
\(
\mathbf D_{\phi_2^c}(\mathbf E_{\phi_1^c}(\boldsymbol G_{c,i}))
\in \mathbb R^{L_c\times 3}
\)
denote the decoded genotype logits for individual \(i\), and let \(p_{cijg}\) be the corresponding softmax probability for genotype class \(g\in\{0,1,2\}\). Let \(\mathcal B\) denote a minibatch. The primary reconstruction term is SNP-wise cross-entropy,
\[
\mathcal L_{\mathrm{CE}}^{(c)}
=
-\frac{1}{|\mathcal B|L_c}
\sum_{i\in\mathcal B}
\sum_{j=1}^{L_c}
\log p_{cij,G_{cij}} .
\]

We also train the autoencoder with auxiliary terms that target population-genetic structure and latent utility for conditional generation:
\[
\mathcal L_{\mathrm{AE}}^{(c)}
=
\mathcal L_{\mathrm{CE}}^{(c)}
+
\lambda_{\mathrm{cond}}\mathcal L_{\mathrm{cond}}^{(c)}
+
\lambda_{\mathrm{MAF}}\mathcal L_{\mathrm{MAF}}^{(c)}
+
\lambda_{\mathrm{LD}}\mathcal L_{\mathrm{LD}}^{(c)}
+
\lambda_{\mathrm{reg}}\mathcal L_{\mathrm{reg}}^{(c)} .
\]
The MAF term compares the empirical allele frequency to the reconstructed expected allele frequency,
\[
\mathcal L_{\mathrm{MAF}}^{(c)}
=
\frac{1}{L_c}
\sum_{j=1}^{L_c}
\left|
\frac{1}{2|\mathcal B|}\sum_{i\in\mathcal B}\widehat G_{cij}
-
\frac{1}{2|\mathcal B|}\sum_{i\in\mathcal B}G_{cij}
\right|,
\]
where \(\widehat G_{cij}=\sum_{g=0}^2 g\,p_{cijg}\). The LD term computes local correlation matrices over non-overlapping SNP windows \(\mathcal W_c\) and penalizes their Frobenius discrepancy:
\[
\mathcal L_{\mathrm{LD}}^{(c)}
=
\frac{1}{|\mathcal W_c|}
\sum_{W\in\mathcal W_c}
\left\|
\mathrm{Corr}(\widehat{\boldsymbol G}_{c,\mathcal B,W})
-
\mathrm{Corr}(\boldsymbol G_{c,\mathcal B,W})
\right\|_F .
\]
Then the condition-prediction term mean-pools latent tokens,
\(
\bar{\boldsymbol z}_{c,i}
=
\frac{1}{T_c}\sum_{t=1}^{T_c}\boldsymbol Z_{c,it},
\)
and predicts the conditioning vector with a two-layer MLP \(h_{\mathrm{cond}}\):
\[
\mathcal L_{\mathrm{cond}}^{(c)}
=
\frac{1}{|\mathcal B|}
\sum_{i\in\mathcal B}
\left\|
h_{\mathrm{cond}}(\bar{\boldsymbol z}_{c,i})-\boldsymbol c_i
\right\|_2^2
\]
to encourage chromosome latents to retain ancestry- and liability-relevant information rather than only local reconstruction information.

Finally, the regularization term \(\mathcal L_{\mathrm{reg}}^{(c)}\) is used only in the final stage of training. It contains latent smoothness, robustness, and stability penalties:
\[
\mathcal L_{\mathrm{reg}}^{(c)}
=
\lambda_{\mathrm{TV}}\mathcal L_{\mathrm{TV}}^{(c)}
+
\lambda_{\mathrm{robust}}\mathcal L_{\mathrm{robust}}^{(c)}
+
\lambda_{\mathrm{stable}}\mathcal L_{\mathrm{stable}}^{(c)} .
\]
The total-variation term penalizes adjacent-token variation,
\[
\mathcal L_{\mathrm{TV}}^{(c)}
=
\frac{1}{T_c-1}
\sum_{t=1}^{T_c-1}
\left\|
\boldsymbol Z_{c,i,t+1}-\boldsymbol Z_{c,it}
\right\|_2^2 .
\]
The ``robustness'' term adds Gaussian noise to the latent representation and penalizes changes in decoded expected dosages, while the ``stability'' term adds Gaussian noise to the embedded input representation and penalizes the resulting latent deviation. These penalties are activated only in the final \(25\%\) of autoencoder training.

\subsubsection{Training hyperparameters}

Each chromosome-specific autoencoder is trained independently using the same objective form and chromosome-specific architecture from Table~\ref{tab:ae_chrom_configs}. Table~\ref{tab:ae_hyperparams} lists the hyperparameters required to reproduce the autoencoder training runs.

\begin{table}[t]
\centering
\caption{Autoencoder training hyperparameters.}
\label{tab:ae_hyperparams}
\small
\setlength{\tabcolsep}{5pt}
\renewcommand{\arraystretch}{0.95}
\begin{tabular}{ll}
\toprule
Hyperparameter & Value \\
\midrule
Optimizer & AdamW, \(\beta=\) \((0.9,0.99)\) \\
Learning rate & \(2\times 10^{-4}\) \\
Learning-rate schedule & None \\
Batch size & \(128\) \\
Number of epochs & 50 \\
Embedding dimension \(d_e\) & \(8\) \\
Latent token dimension \(d\) & \(256\) \\
Total genome-wide token budget \(\sum_c T_c\) & \(4096\) \\
Condition MLP hidden dimension & \(256\) \\
LD window size & \(128\) \\
Latent regularization penalties active & Final \(25\)\% of training \\
\(\lambda_{\mathrm{cond}}\) & \(1.0\) \\
\(\lambda_{\mathrm{MAF}}\) & \(10^{-3}\) \\
\(\lambda_{\mathrm{LD}}\) & \(10^{-3}\) \\
\(\lambda_{\mathrm{TV}}\) & \(2\times 10^{-2}\) \\
\(\lambda_{\mathrm{robust}}\) & \(100.0\) \\
\(\lambda_{\mathrm{stable}}\) & \(3\times 10^{-1}\) \\
Latent Gaussian noise $\sigma$ (robustness) & \(0.05\) \\
Input-embedding Gaussian noise $\sigma$ (stability) & \(0.03\) \\
Dropout & \(0.0\) \\
Weight decay & \(0.0\) \\
Gradient clipping & \(1.0\) \\
\bottomrule
\end{tabular}
\end{table}

\subsubsection{Regularization ablations}
\label{sec:app-ablateLreg}

\begin{figure}
    \centering
    \includegraphics[width=\linewidth]{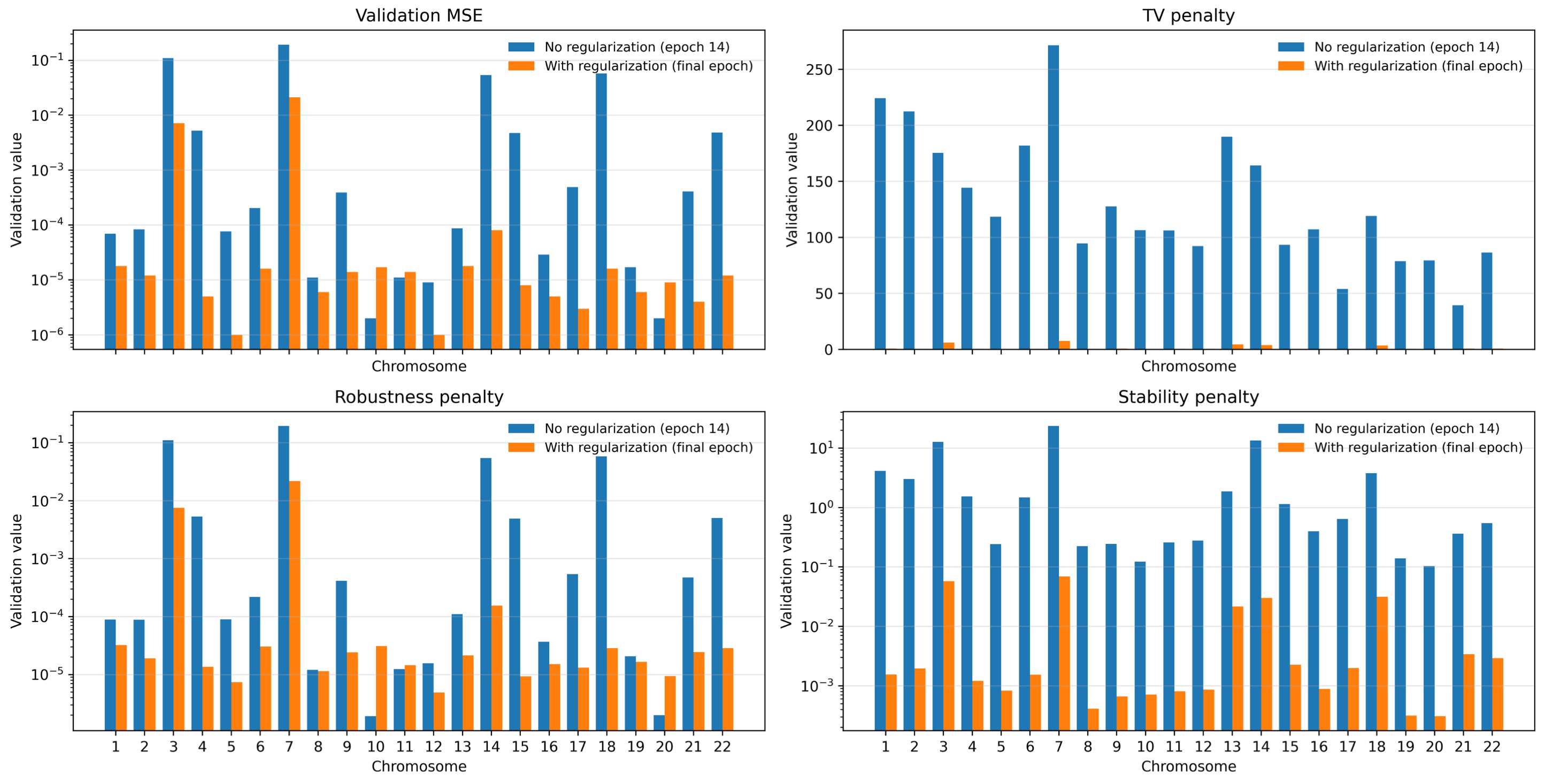}
    \caption{Regularization ablation for chromosome-specific autoencoders. For each chromosome, we compare validation metrics at the last epoch before $\lambda_{\rm reg}>0$, against the final epoch after regularization. Regularization consistently reduces latent total variation, robustness loss, and stability loss, while continuing to improve validation reconstruction MSE.}
    \label{fig:regAblate}
\end{figure}

We ablate the auxiliary loss terms to assess whether they improve reconstruction fidelity and the suitability of the latent space for downstream flow matching. We do this by comparing validation performance at the final epoch before $\lambda_{\rm reg}>0$ to the final epoch of autoencoder training (Figure~\ref{fig:regAblate}). 
We see regularization does not degrade reconstruction accuracy; validation MSE is lower at the final regularized epoch for most chromosomes. The largest effect is on the latent TV penalty, which drops sharply after regularization, suggesting that the token latents become substantially smoother. Robustness and stability penalties also decrease, indicating that the learned latent representation becomes less sensitive to latent and input-embedding perturbations, which we posit provides a better-conditioned representation for downstream flow matching.

\subsection{Latent flow matching model}
\label{app:usit_details}

After training the chromosome-specific autoencoders, we freeze all encoders and decoders and train a generative model over the concatenated genome-wide latent token sequence. For individual \(i\), let \(\boldsymbol Z_i=\{\boldsymbol Z_{1,i},\ldots,\boldsymbol Z_{22,i}\}\in\mathbb R^{T\times d}\) denote the concatenation of chromosome-specific autoencoder latents, where \(T=\sum_c T_c\). We train the flow model with a linear interpolant flow matching objective. For each training latent \({\boldsymbol Z}_1\), we sample
\(
\boldsymbol Z_0\sim \mathcal N(0,I)
\) and 
\(
t\sim \mathrm{Unif}(0,1),
\)
and construct
\(
\boldsymbol Z_t = (1-t)\boldsymbol Z_0 + t{\boldsymbol Z}_1.
\)
The target velocity is
\(
v^\star = {\boldsymbol Z}_1-\boldsymbol Z_0.
\)
We train \(v_\theta(\boldsymbol Z_t,t,\boldsymbol c)\), where \(\boldsymbol c\) is the conditioning profile for conditional models, with velocity MSE:
\[
\mathcal L_{\mathrm{FM}}
=
\mathbb E_{t,\boldsymbol Z_0,{\boldsymbol Z}_1,\boldsymbol c}
\left[
\left\|
v_\theta(\boldsymbol Z_t,t,\boldsymbol c)
-
({\boldsymbol Z}_1-\boldsymbol Z_0)
\right\|_2^2
\right].
\]
The noisy latent sequence \(\boldsymbol Z_t\in\mathbb R^{T\times d}\) is linearly projected to \(\boldsymbol H_t\in\mathbb R^{T\times D_h}\), with \(d=256\) and hidden width \(D_h=1152\) in Table~\ref{tab:flow_training}.
We then add learned positional embeddings and chromosome/source embeddings to the token sequence. To provide chromosome-specific capacity while preserving global self-attention, we use lightweight source-aware adapters indexed by chromosome. Specifically, for token \(t\) from source \(s(t)\), the adapter computes
\[
\Delta \boldsymbol H_t
=
A_{s(t)}\mathrm{LN}(\boldsymbol H_t)+\boldsymbol b_{s(t)},
\]
and updates
\(
\boldsymbol H_t
\leftarrow
\boldsymbol H_t
+
g_{s(t)}\Delta \boldsymbol H_t ,
\)
where \(A_s\), \(\boldsymbol b_s\), and \(g_s\) are source-specific parameters. The adapter gate \(g_s\) is initialized to zero, so the source adapter is inactive at initialization and learned gradually.

Time is embedded with Fourier features followed by an MLP. For conditional models, the condition vector is embedded with a separate MLP. These embeddings are used both as special tokens prepended to the latent sequence and as the conditioning signal for DiT-style adaLN-Zero modulation \cite{peebles2023scalable} inside each transformer block. Conditional training supports classifier-free guidance by replacing the condition embedding with a learned null embedding with probability \(p_{\mathrm{drop}}\); guidance-scale ablations are reported in Appendix~\ref{app:cfg}. Training and sampling hyperparameters are summarized in Table~\ref{tab:flow_training}.

\subsubsection{USiT architecture}

\begin{figure}[h]
    \centering
    \includegraphics[width=0.2\linewidth]{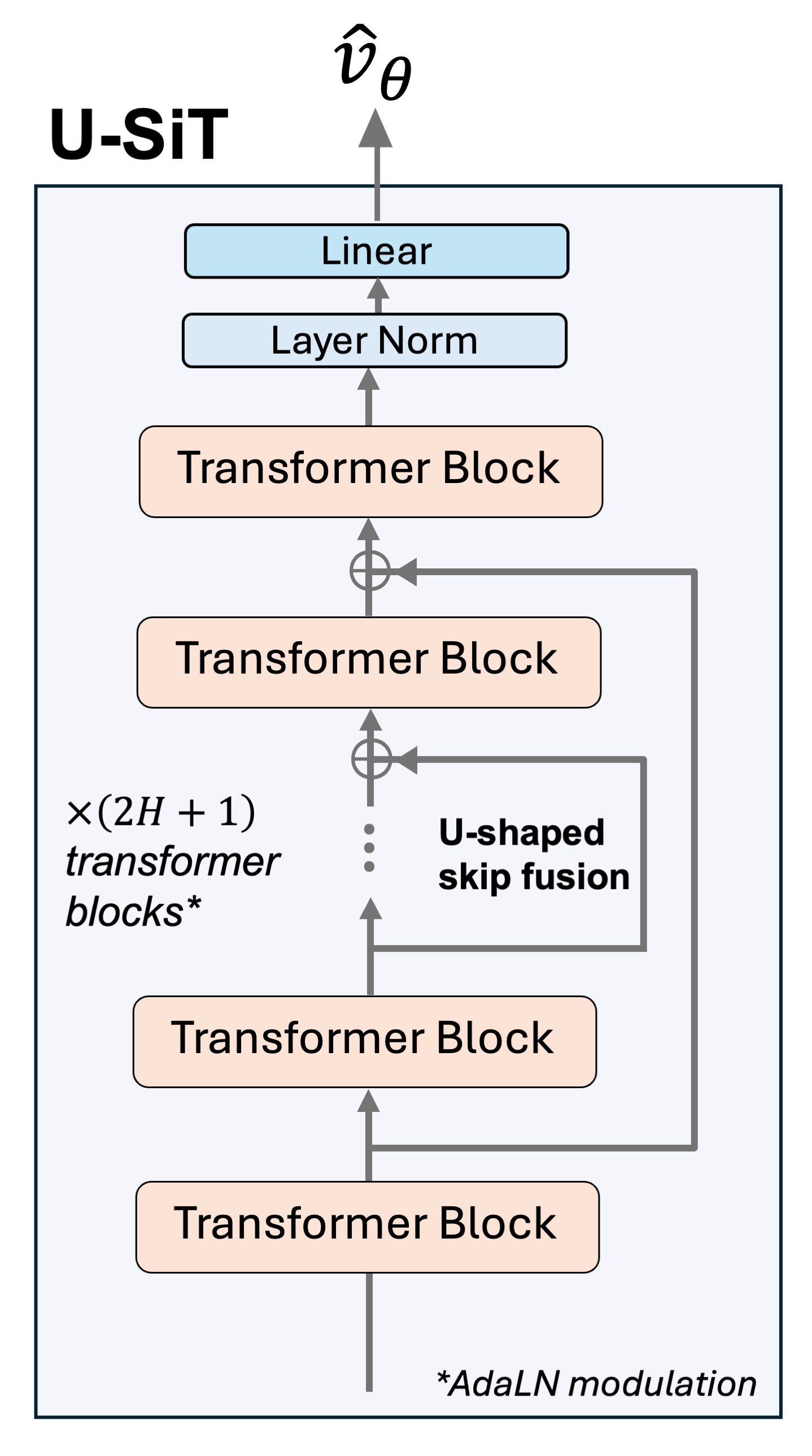}
    \caption{USiT block schematic demonstrating long-range skip connections.}
    \label{fig:usit_block}
\end{figure}

We parameterize the velocity field with a U-shaped Scalable Interpolant Transformer (USiT), shown in Figure~\ref{fig:usit_block}. USiT applies full multi-head self-attention over the latent-token sequence and special tokens, with DiT-style adaLN-Zero modulation from the time and condition embeddings. The backbone contains an odd number of transformer layers, \(2H+1\), organized as \(H\) input blocks, one middle block, and \(H\) output blocks. Activations from the input blocks are passed to the corresponding output blocks through U-shaped skip connections, fused by concatenation followed by a linear projection. This preserves token-level information while still allowing genome-wide mixing through full self-attention. Each block is a pre-norm transformer block with multi-head self-attention and a GELU MLP with hidden width \(rD_h\), where \(r\) is the MLP ratio. In conditional models, both branches use adaLN-Zero modulation. After the USiT backbone, we discard the special time and condition tokens and linearly project the remaining \(T\) token outputs back to the autoencoder latent dimension, yielding \(v_\theta(\boldsymbol Z_t,t,\boldsymbol c)\in\mathbb R^{T\times d}\).

\begin{table}[t]
\centering
\caption{Latent flow matching training and sampling configuration.}
\label{tab:flow_training}
\small
\setlength{\tabcolsep}{5pt}
\renewcommand{\arraystretch}{0.95}
\begin{tabular}{ll}
\toprule
Parameter & Value \\
\midrule
Hardware & \(4\times\) NVIDIA H200 GPUs \\
Training budget & 72 Hours \\
Input latent shape & \((4096,256)\) \\
Backbone & USiT \\
Transformer hidden width & \(1152\) \\
Transformer layers & \(29\) \\
Attention heads & \(16\) \\
MLP ratio & \(4\) \\
Dropout & \(0.0\) \\
Conditioning & Condition token + adaLN-Zero \\
Optimizer & AdamW, \(\beta=(0.9,0.99)\) \\
Learning-rate schedule & warmup + cosine decay \\
Peak learning rate & \(10^{-5}\) \\
Weight decay & \(10^{-5}\) \\
Batch size & \(8\) per GPU; \(32\) global  \\
Training steps & 91,427 \\
Gradient clipping & \(1.0\) \\
EMA decay & \(0.999\) \\
CFG dropout & \(0.2\) \\
Sampler & DPM-Solver++ \\
Sampling steps & 25 \\
Guidance scale & \(1.0\) \\
\bottomrule
\end{tabular}
\end{table}

\subsubsection{Generation}

At generation time, we sample
\(
\boldsymbol Z_0\sim \mathcal N(0,\boldsymbol I)
\), specify a condition vector \(\boldsymbol c\), and integrate the learned ODE
\(
d\boldsymbol Z_t/dt
=
v_\theta(\boldsymbol Z_t,t,\boldsymbol c)
\)
from \(t=0\) to \(t=1\) using DPM-Solver++ \cite{lu2025dpm} for numerical integration. Main results use standard conditional sampling, equivalent to classifier-free guidance \cite{ho2022classifier} with a scale of \(1\); Appendix~\ref{app:cfg} evaluates other scales. After generation, the latent sequence is then partitioned into chromosome-specific blocks. Each block is decoded with the corresponding frozen chromosome decoder to obtain genotype logits, which are converted to hard genotype calls by argmax over the three dosage classes.

\section{Post-training refinement}
\label{app:post_generation_refinement}

For the multi-ancestry HELICS model, we apply a post-training refinement procedure before final benchmarking, which includes token-wise spectral projection before decoding and genotype-space PC alignment followed by deterministic rounding after decoding. HELICS-EUR does not use these refinements. Spectral statistics are estimated from training latents. Genotype standardization, PCA bases, and alignment parameters are estimated once using training genotypes and a fixed generated calibration sample, then frozen. No CCDG or held-out evaluation-cohort information is used to fit these transformations. This closed-form decoder-side refinement is motivated by refinement strategies in compressed latent generative models \cite{chen2024deep, yu2024image, chen2025masked}. For example, DC-AE architectures demonstrated that high-compression autoencoders benefit from including refinement steps to tune middle layers (adapt the latent space) and decoder head layers (improve local reconstruction without changing the full latent representation) \cite{chen2024deep}. We apply the same principle here using a token-wise latent refinement before decoding and a genotype-space PC alignment after decoding, instead of additional training.

\subsection{Token-wise latent spectral truncation}
\label{sec:app-spectralTrunc}
Let \(\widehat{\boldsymbol Z}\in\mathbb R^{n_g\times T\times d}\) denote generated latent tokens in the autoencoder latent space. For each token index \(t\), we fit PCA to the corresponding training latents \(\boldsymbol Z_{\mathrm{train},t}\in\mathbb R^{n_o\times d}\), obtaining mean \(\boldsymbol\mu_t\), components \(\boldsymbol U_t\), and explained variances \(\boldsymbol\lambda_t\). We then project generated token vectors onto the smallest token-specific principal subspace explaining \(99\%\) of training variance:
\[
k_t
=
\min\left\{
k:
\frac{\sum_{\ell=1}^{k}\lambda_{t\ell}}
{\sum_{\ell}\lambda_{t\ell}}
\ge 0.99
\right\}.
\]
The corrected generated token is
\[
\widehat{\boldsymbol Z}^{\mathrm{spec}}_{:,t,:}
=
\boldsymbol\mu_t
+
\left[
(\widehat{\boldsymbol Z}_{:,t,:}-\boldsymbol\mu_t)
\boldsymbol U_{t,1:k_t}^{\top}
\right]
\boldsymbol U_{t,1:k_t}.
\]
This projection removes generated variation along low-variance token directions that are poorly supported by the training latent distribution, while preserving the dominant token-wise latent structure. The \(99\%\) cutoff is prespecified. All token means, bases, eigenvalues, and retained dimensions are estimated from training latents and fixed.

\subsection{Genotype-space PC alignment}

After spectral truncation, we decode a fixed generated calibration sample, taking the argmax over genotype logits to obtain synthetic genotype calls \(\widetilde{\boldsymbol G}_c\). We estimate genotype-PC alignment from this calibration sample and training genotypes, then retain the transformation for subsequent generations. For chromosome \(c\), let
\(
\boldsymbol G_c\in\mathbb R^{n\times L_c},
\widetilde{\boldsymbol G}_c\in\mathbb R^{n\times L_c}
\)
denote standardized training and generated calibration genotypes over the same \(L_c\) variants; we use $n$ training and generated individuals for this alignment. All means, scales, and PCA loadings used in this correction are retained and frozen. We compute separate \(k\)-dimensional PCA bases for the original and generated genotypes such that 
\(
\boldsymbol G_c \approx \boldsymbol S_c\boldsymbol W_c,
\) and
\(
\widetilde{\boldsymbol G}_c \approx \widetilde{\boldsymbol S}_c\widetilde{\boldsymbol W}_c,
\)
where \(\boldsymbol W_c,\widetilde{\boldsymbol W}_c\in\mathbb R^{k\times L_c}\) are loading matrices and \(\boldsymbol S_c,\widetilde{\boldsymbol S}_c\) are PC score matrices. We estimate an orthogonal Procrustes rotation aligning generated PC axes to the original PC axes:
\[
\boldsymbol R_c^\star
=
\arg\min_{\boldsymbol R\in O(k)}
\left\|
\boldsymbol W_c-\boldsymbol R\widetilde{\boldsymbol W}_c
\right\|_F^2.
\]
The Procrustes problem has a known closed-form solution \cite{schonemann1966generalized}; if
\(
\boldsymbol W_c\widetilde{\boldsymbol W}_c^\top
=
\boldsymbol U\boldsymbol\Sigma\boldsymbol V^\top,
\)
then 
\(
\boldsymbol R_c^\star=\boldsymbol U\boldsymbol V^\top.
\)
For a subsequent generated sample, we use the fixed calibration standardization and loading matrix \(\widetilde{\boldsymbol W}_c\) to obtain its generated-PC scores; PCA and alignment are not refitted to the new generation batch. The generated scores are expressed in the original PC basis by
\(
\widetilde{\boldsymbol S}_c^{\mathrm{corr}}
=
\widetilde{\boldsymbol S}_c{\boldsymbol R_c^\star}^{\top},
\)
and hence the aligned standardized genotypes are reconstructed as
\(
\widetilde{\boldsymbol G}_c^{\mathrm{corr}}
=
\widetilde{\boldsymbol S}_c^{\mathrm{corr}}\boldsymbol W_c.
\)
Note that the number of retained PCs \(k\) is chosen during the post-training refinement procedure for each chromosome to maximize overall fidelity across MAF, LD, and PC-alignment metrics. 
After reversing the genotype standardization, we clip values to the valid dosage range \([0,2]\) and discretize to hard genotype calls, placing generated samples in the same chromosome-specific genetic coordinate system as the training cohort while leaving the trained autoencoders and flow model fixed. Recognize that since this correction reduces to standardization followed by a low-rank matrix multiplication, it can equivalently be viewed as a fixed final linear correction layer after the decoder, making explicit the connection between our closed-form refinement and the decoder-head refinement used in DC-AE-style adaptation.

\subsection{Ablation of post-training refinement}
\label{app:post_generation_refinement_ablation}

In this ablation study, we show the need for both components of the post-training refinement procedure for the multi-ancestry HELICS model. 

\begin{figure}[h]
    \centering
    \includegraphics[width=\linewidth]{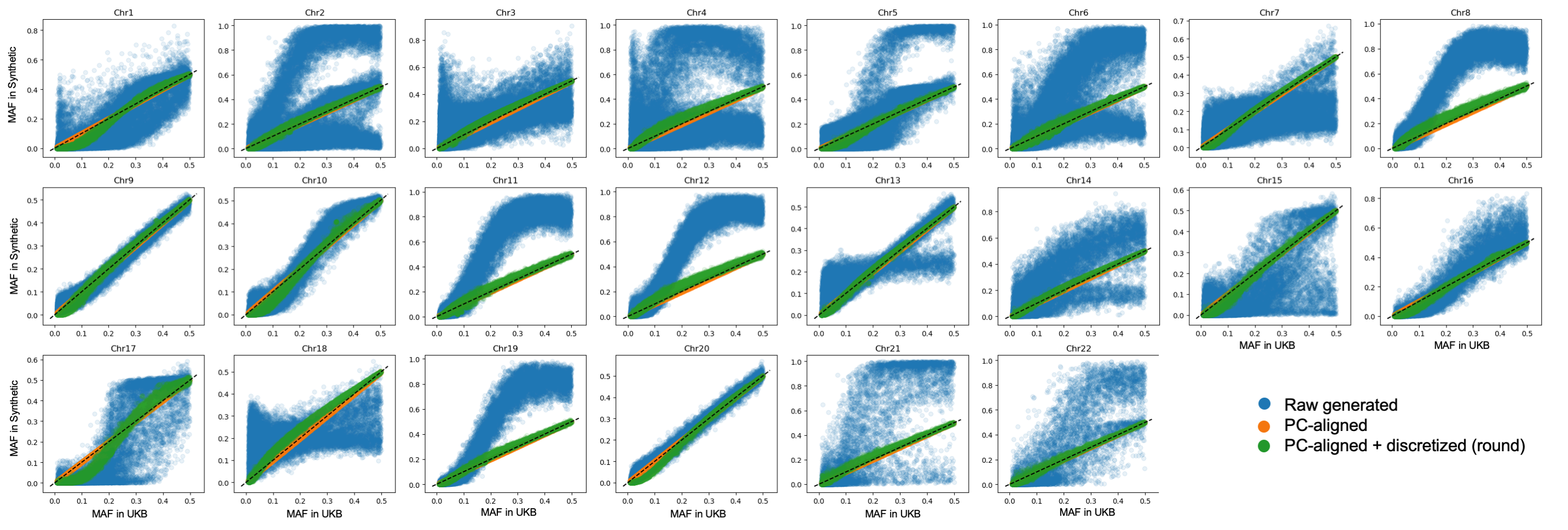}
    \caption{
    Effect of genotype-space PC alignment on chromosome-wise MAF fidelity. 
    Each panel compares UKB reference MAF to synthetic MAF for one chromosome. 
    Raw generated genotypes show distortions in marginal allele frequencies, consistent with a global population-structure misalignment. The PC-alignment correction substantially improves marginal agreement across chromosomes, and the final discretized calls preserve most of this improvement.
    }
    \label{fig:postgen_maf_chr}
\end{figure}
\paragraph{Effect of genotype-space PC alignment.}
Figure~\ref{fig:postgen_maf_chr} compares chromosome-wise MAF agreement before PC alignment, after PC alignment, and after final discretization. PC alignment substantially improves marginal agreement across chromosomes, consistent with correcting a low-rank discrepancy in population structure. This correction is especially important in the multi-ancestry setting, where the generated genotype space can appear rotated relative to the training cohort. We hypothesize that this arises because the latent representation has many more scalar degrees of freedom than the underlying low-dimensional population structure, leaving the flow model room to represent ancestry geometry in a rotated latent coordinate system. 

\begin{figure}[h]
    \centering
    \includegraphics[width=\linewidth]{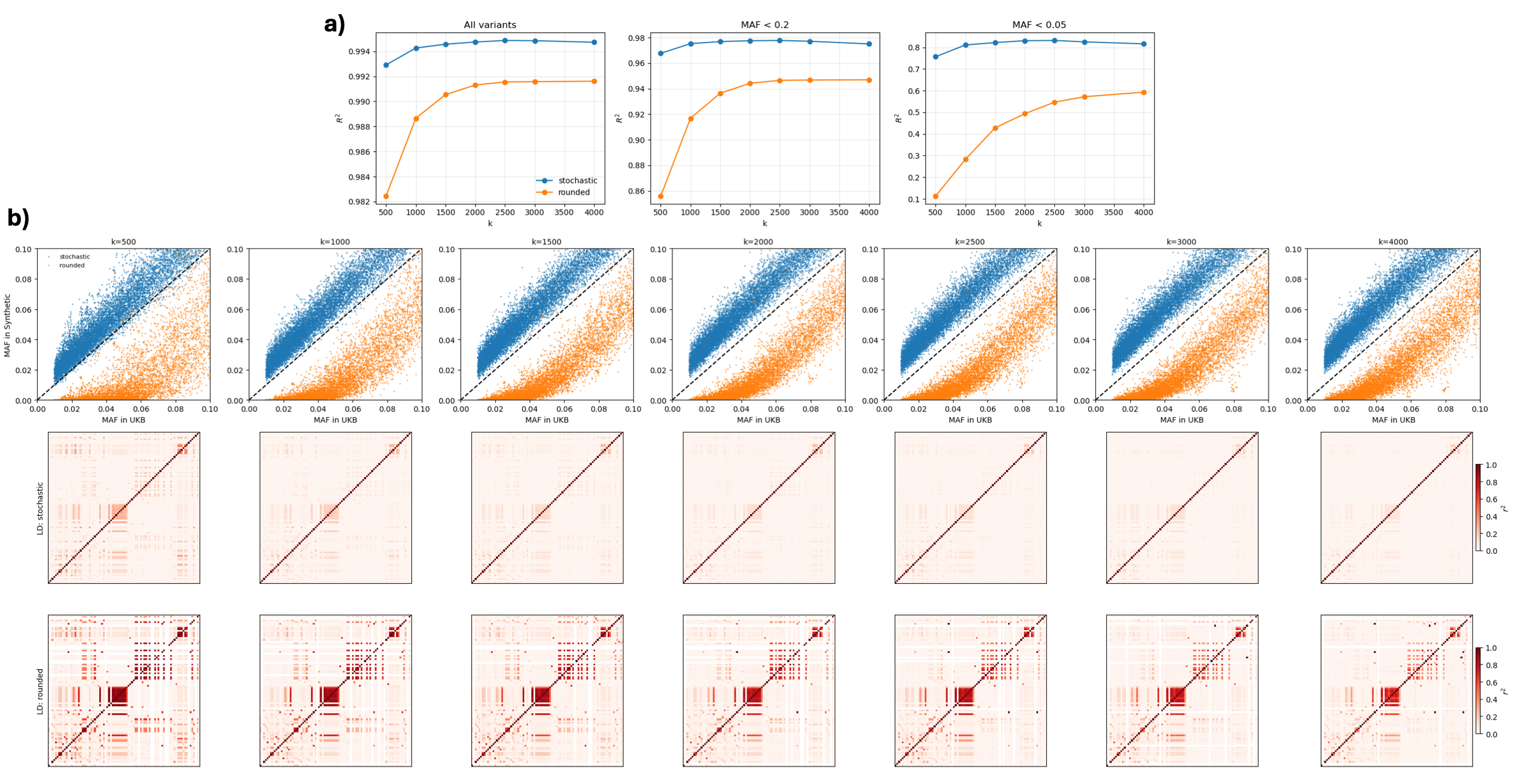}
    \caption{
    Effect of retained PC dimension \(k\) and genotype discretization strategy after PC alignment. (a) MAF $R^2$ correspondence across several values of $k$ for all variants, variants with $0.01 < MAF < 0.2$, and $0.01 < MAF < 0.05$. (b) Top row: low-frequency MAF (\(<0.1\)) agreement under stochastic genotype calling versus deterministic rounding. 
    Representative LD matrices are also shown after stochastic calling (middle row) and deterministic rounding (bottom row). 
    }
    \label{fig:postgen_rounding}
\end{figure}
\paragraph{Choice of discretization.}
After PC alignment, continuous corrected dosages must be converted back to discrete genotype calls. Figure~\ref{fig:postgen_rounding} compares deterministic clipping and rounding with stochastic genotype calling across different values of the retained PC dimension \(k\). Here, stochastic genotype calling samples a hard genotype \(g\in\{0,1,2\}\) with probabilities determined by the corrected continuous dosage's relative proximity to the three valid dosage states, rather than rounding to the nearest integer deterministically. Stochastic rounding improves low-frequency marginal alignment, particularly for low-MAF variants, but it introduces additional noise into local correlation structure and degrades LD fidelity. In contrast, deterministic clipping and hard discretization preserve LD structure more consistently. We therefore use deterministic clipping to \([0,2]\) followed by hard genotype calls in the final pipeline. As \(k\) increases, MAF fidelity improves and plateaus whereas LD degrades, motivating selection of \(k\) within the post-training refinement procedure and freezing it before held-out evaluation.

\begin{figure}[h]
    \centering
    \includegraphics[width=\linewidth]{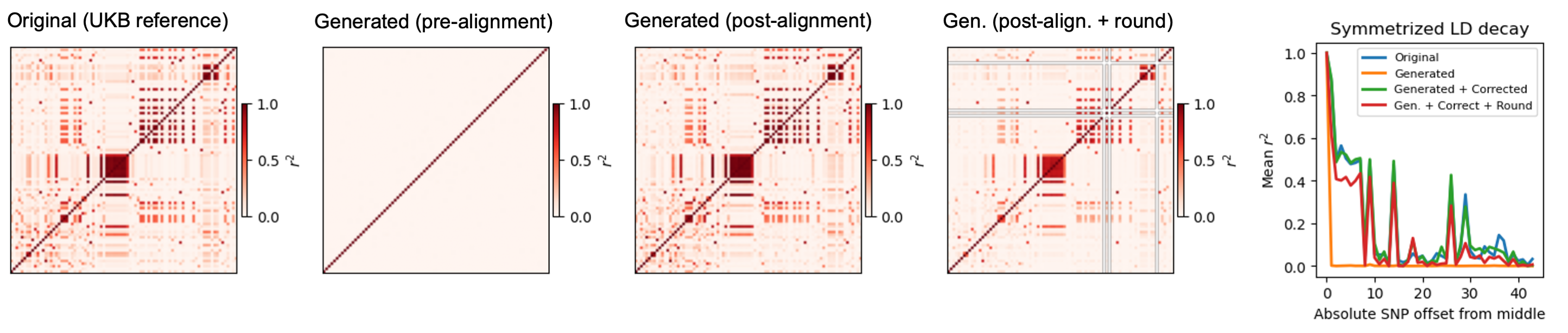}
    \caption{
    Effect of PC alignment on local LD structure. 
    A representative LD block (chr15:24{,}321{,}414--24{,}848{,}545) is shown for the UKB reference, raw generated genotypes, PC-aligned generated genotypes, and PC-aligned genotypes after final rounding. 
    A symmetrized LD-decay curve, computed from the center of the LD block, is also shown.
    }
    \label{fig:postgen_ld_block}
\end{figure}
\paragraph{Effect on LD structure.}
Figure~\ref{fig:postgen_ld_block} illustrates the effect of refinement on one representative LD block. Raw generated genotypes show weak local LD structure in this block, while the PC-corrected genotypes recover the major block pattern observed in the original cohort. Final rounding slightly attenuates the recovered LD but preserves substantially more structure than the uncorrected generated data. The symmetrized LD-decay curve shows the same qualitative trend; PC correction restores short- and medium-range LD structure, whereas discretization introduces a modest loss relative to the continuous corrected dosages.

\begin{figure}[h]
    \centering
    \includegraphics[width=\linewidth]{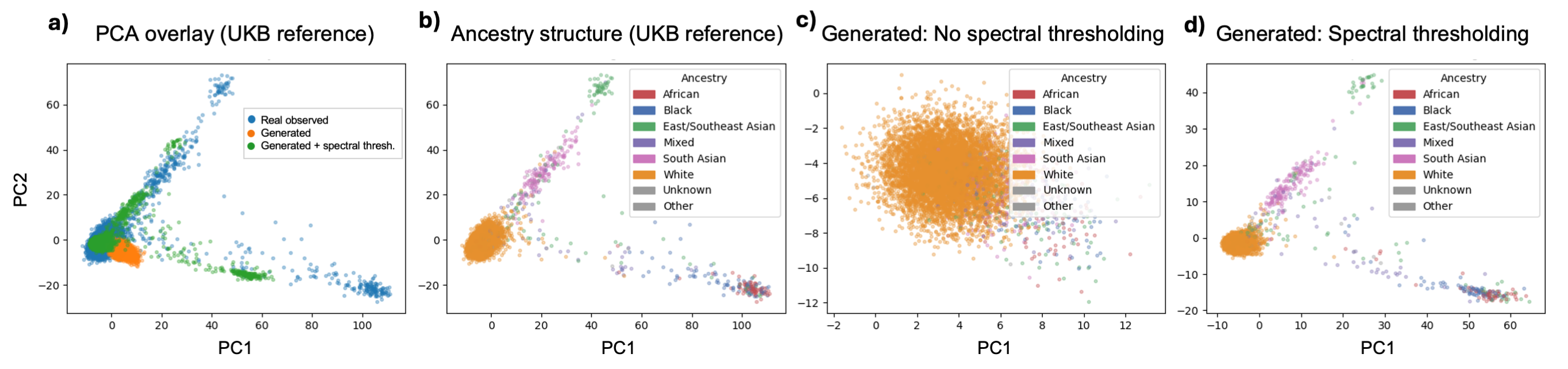}
    \caption{Effect of token-wise spectral truncation on multi-ancestry population structure. 
    (a) Overlay of generated samples on UKB genotype-PC space. (b) UKB ancestry structure based on chromosome 15 variants. (c) PC-projected generated samples without spectral thresholding; generated samples collapse in PC space and ancestry groups are poorly separated.  (d) Generated samples after spectral thresholding. 
    }
    \label{fig:postgen_spectral}
\end{figure}
\paragraph{Effect of token-wise spectral truncation.}
Figure~\ref{fig:postgen_spectral} ablates the latent spectral correction. Without spectral truncation, generated samples collapse in genotype-PC space and fail to separate ancestry groups, even after decoding. Applying token-wise spectral thresholding before decoding restores the dominant ancestry geometry; namely, generated samples show the expected genetic ancestry structure and self-reported ancestry groups separate. This indicates that low-variance token directions in the generated latent distribution can decode into unstable population-structure artifacts. Removing these unsupported directions before decoding is therefore necessary for reliable multi-ancestry generation.

Together, these ablations demonstrate that post-training refinement is a necessary component for multi-ancestry HELICS. Spectral truncation stabilizes generated latents before decoding, while genotype-space PC alignment corrects broad low-rank population-structure discrepancies in decoded genotypes. However, while the refinement improves PC alignment and recovers LD structure and MAF agreement, the final discretization step reduces low-frequency MAF agreement due to the discretization procedure.

\section{Conditioning construction and guided sampling}
\label{app:conditions}

\subsection{Condition vector definition}

HELICS conditions generation on a profile \(\boldsymbol c_i\) containing continuous ancestry coordinates and discretized genetic-liability summaries. Specifically, \(\boldsymbol c_i\) includes the first 20 genotype principal components (PCs) and trait-specific PRS bins. Genotype PCs are used rather than self-reported ancestry labels as they provide a continuous coordinate system for ancestry, admixture, and fine-scale population structure; we also conveniently circumvent any errors or missingness in self-reported ancestry. For each trait \(k\), we compute a published polygenic risk score
\[
\mathrm{PRS}_{ik}
=
\sum_{j\in\mathcal S_k}
\beta_{jk}G_{ij},
\]
where \(\mathcal S_k\) is the set of matched score variants, \(\beta_{jk}\) is the published effect weight, and \(G_{ij}\in\{0,1,2\}\) is the dosage of the reported effect allele. A high PRS bin is evidence of high genetic liability for a trait.

We selected traits spanning anthropometric, cardiometabolic, autoimmune, psychiatric, renal, liver-enzyme, and cancer-related genetic liabilities. PRS weights were obtained from the PGS Catalog using the accessions listed in Table~\ref{tab:prs_sources}. These scores were used only to construct individual-level conditioning summaries and to evaluate perturbation responses; HELICS is not given variant-level PRS weights as model inputs.

\begin{table}[t]
\centering
\caption{Polygenic scores used to construct trait-conditioning variables.}
\label{tab:prs_sources}

\setlength{\tabcolsep}{5pt}
\renewcommand{\arraystretch}{0.95}
\begin{tabular}{lc}
\toprule
Trait & PGS Catalog accession \\
\midrule
Standing height & PGS002332 \\
Coronary artery disease & PGS003725 \\
Alzheimer's disease (dementia) & PGS003958 \\
Major depression &  PGS003333 \\
Body mass index &  PGS005198 \\
High-density lipoprotein cholesterol & PGS004914 \\
Chronic kidney disease (eGFR) & PGS000883 \\
Gamma-glutamyl transferase & PGS003543 \\
Alanine aminotransferase  & PGS000668 \\
Low-density lipoprotein cholesterol  & PGS000893 \\
Total cholesterol & PGS003143 \\
Apolipoprotein A  & PGS002101 \\
Apolipoprotein B  & PGS002102 \\
Hemoglobin A1c &  PGS004703 \\
Type 1 diabetes  & PGS002025 \\
Type 2 diabetes & PGS000330 \\
Lung cancer & PGS000395 \\
Rheumatoid arthritis & PGS002088 \\
Hypertension & PGS002047 \\
\bottomrule
\end{tabular}
\end{table}

For each PRS, score variants were harmonized to the UKB genotype representation before scoring. Variants were matched by genomic position and allele pair after build harmonization. Effect alleles were aligned to the UKB A1/A2 convention using allele swaps and strand flips and removing ambiguous variants. Scores were standardized using the training cohort:
\[
\widetilde{\mathrm{PRS}}_{ik}
=
\frac{\mathrm{PRS}_{ik}-\mu_k^{\mathrm{train}}}
{\sigma_k^{\mathrm{train}}},
\]
where \(\mu_k^{\mathrm{train}}\) and \(\sigma_k^{\mathrm{train}}\) are the training-set mean and standard deviation for trait \(k\). We then discretized each standardized PRS into decile bins using training-set quantile cutoffs. The resulting bin index, together with the top 20 genotype PCs, forms the conditioning profile \(\boldsymbol c_i\). 

For conditional generation experiments, we condition on profiles sampled from held-out validation individuals. This avoids learning a separate model for the joint distribution of ancestry PCs and PRS bins, though such a model is straightforward to train with any flexible learner.

Let \(\mathcal S_{20}\) contain the 20 variants with the largest absolute PRS contributions \(|\beta_j\Delta G_j|\). The weighted Top-20 metric in Table~\ref{tab:conditional_control} is
\[
\frac{\sum_{j\in\mathcal S_{20}}|\beta_j\Delta G_j|\,
\mathbf 1\{\beta_j\Delta G_j>0\}}
{\sum_{j\in\mathcal S_{20}}|\beta_j\Delta G_j|}.
\]
Ancestry-stratified results in Table~\ref{tab:conditional_control_ancestry} report the unweighted fraction satisfying \(\beta_j\Delta G_j>0\), with dosage shifts and rankings recomputed within each ancestry.

\subsection{Validation with external polygenic scores}
\label{app:external_hdl}

Evaluation using the same PRS that defines conditioning measures consistency with the supplied target. To assess generalization beyond the conditioning score, we rescored generated genomes using PGS002960, a separate HDL score not used for conditioning; HDL conditioning used PGS004914 deciles (Table~\ref{tab:prs_sources}). We measured the correlation between continuous PGS002960 scores in generated genomes and their paired held-out real EUR genomes. We additionally measured the correlation between PGS002960 and PGS004914 within these real genomes. Scores were computed as effect-allele dosage-weight sums after matching variants by chromosome, position, and effect allele, with missing dosages mean-imputed within each dataset. We repeated the analysis excluding external-score variants in LD with any conditioning-score variants (\(r^2\geq0.8\)), including overlapping variants. 

Positive generated-observed correspondence persists after LD filtering (Table~\ref{tab:external_hdl}), supporting recovery of external HDL-score variation under validation-profile conditioning. These results extend validation beyond consistency with the supplied conditioning score.

\begin{table}[h]
\centering
\caption{External HDL-score validation. Generated-observed \(r\) correlates continuous PGS002960 scores between generated genomes and their paired real genomes under validation-profile conditioning. Real cross-score \(r\) summarizes the association between PGS002960 and PGS004914 within the real cohort. The second row excludes external-score variants with \(r^2\geq0.8\) to any matched conditioning-score variant on the same chromosome.}
\label{tab:external_hdl}
\small
\setlength{\tabcolsep}{4pt}
\begin{tabular}{lccc}
\toprule
\textbf{PGS002960 (external) variant set} & \textbf{Variants scored} & \textbf{Generated-observed} \(r\) & \textbf{Real cross-score} \(r\) \\
\midrule
All matched variants & 2{,}627 & 0.189 & 0.300 \\
After LD exclusions & 2{,}608 & 0.135 & 0.194 \\
\bottomrule
\end{tabular}
\end{table}

\subsection{Classifier-free guidance experiments}
\label{app:cfg}

\begin{figure}
    \centering
    \includegraphics[width=\linewidth]{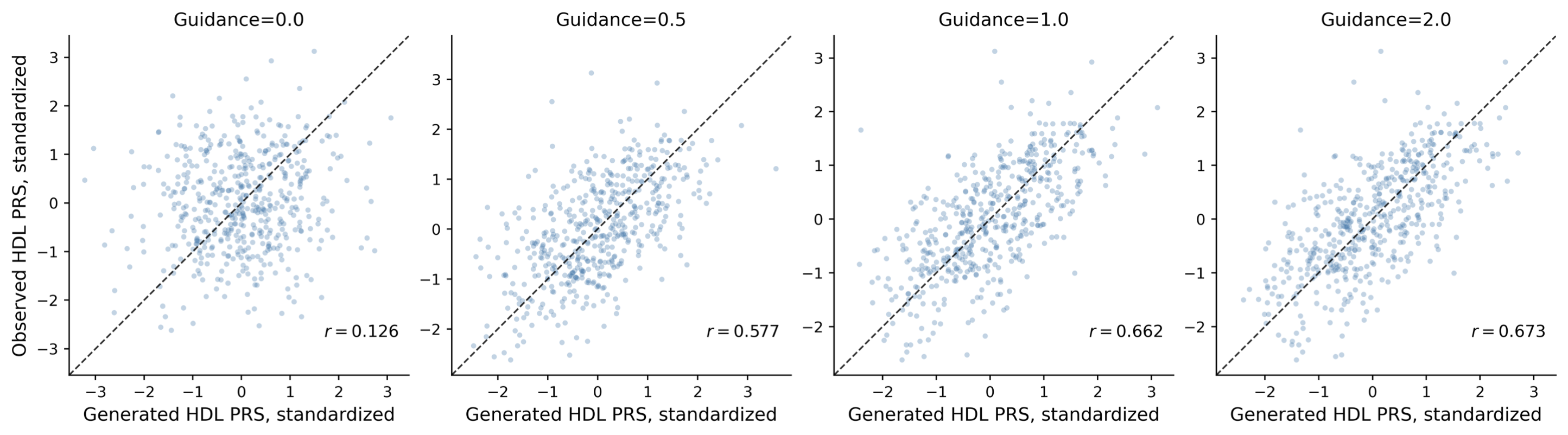}
    \caption{Classifier-free guidance strengthens PRS control in HELICS. 
    Each panel compares standardized generated HDL PRS to standardized observed HDL PRS for held-out validation conditioning profiles under a different guidance scale \(s\); see Figure~\ref{fig:prs_hdl_schematic} for an illustration of this procedure. 
    As \(s\) increases from \(0\) to \(2\), generated HDL PRS becomes more strongly correlated with the conditioning profile, indicating tighter conditional control.
    }
    \label{fig:cfg_hdl}
\end{figure}

The conditional flow model supports classifier-free guidance by replacing the condition embedding with a learned null condition during training with probability \(p_{\mathrm{drop}}\). At sampling time, conditional and unconditional velocity predictions can be combined as
\[
v_{\mathrm{CFG}}
=
v_{\mathrm{uncond}}
+
s\left(v_{\mathrm{cond}}-v_{\mathrm{uncond}}\right),
\]
where \(s\) is the guidance scale. Main results use \(s=1.0\). While this mechanism is available in HELICS, we do not use classifier-free guidance for the main experiments in this work; all reported results use standard conditional sampling. We include Figure~\ref{fig:cfg_hdl} to demonstrate that guidance behaves as expected: increasing \(s\) strengthens correspondence between generated and observed HDL PRS, while \(s=0\) removes most condition-specific control.

\section{Existing synthetic genome generators}
\label{app:related-work}

We survey two broad classes of synthetic genotype generators: reference-panel simulators that rely on observed input haplotypes, and learned generators trained on fixed SNP panels. HAPGEN2 \cite{su2011hapgen2} is a classical haplotype-based simulator that generates synthetic genomes by simulating recombination or resampling haplotypic structure from the input reference panel. It was designed for simulating case-control datasets at SNP markers in particular, including settings with multiple disease SNPs on a chromosome. Sim1000G \cite{dimitromanolakis2019sim1000g} estimates regional allele frequency and a dense covariance matrix and generates genotypes by thresholding multivariate Gaussian samples; this provides strong regional frequency fidelity observed in Table~\ref{tab:main_benchmark}. Dense covariance storage scales as \(O(L^2)\) and factorization as \(O(L^3)\). A \(604{,}243\times604{,}243\) double-precision covariance matrix alone requires approximately \(2.9\) TB, before factorization and other arrays. While inflexible and typically not scalable, these approaches are strong baselines for local haplotype realism as they preserve local correlation structure present in the reference panel. These approaches are best interpreted as reference-panel simulators as opposed to flexible genome-wide generative models. They preserve AF and local LD when the reference panel is well matched but are not suited for learning new genotype distributions from biobank-scale cohorts or for conditioning on genetic-liability profiles.

HAPNEST \cite{wharrie2023hapnest} is a more recent reference-based simulator. Developed to generate large synthetic genotype and phenotype cohorts, it improves the scalability of haplotype-based simulation and was designed to preserve data fidelity while reducing close genetic relatedness to the reference panel. Nevertheless, HAPNEST remains tied to phased reference haplotype structure, so while it is a strong large-scale baseline, its reliance on phased genotypes and reference-panel based haplotype structure are serious limitations.

Recent work \cite{yelmenGAN} introduced deep generative models for artificial human genomes using generative adversarial networks (GANs) and restricted Boltzmann machines (RBMs). These models were trained on phased haplotypes and evaluated on a small \(10{,}000\)-SNP chromosome 15 panel. While limited in terms of genomic scale, this work demonstrated that neural generators can in fact reproduce important properties of real genomic data, including AFs, LD, pairwise haplotype distances, and population structure. Nevertheless, dense neural generators trained directly on SNP haplotypes become difficult to scale as the number of variants grows. Later work used convolutional WGAN and conditional RBM models for larger genomic segments, including chromosome 1-scale panels \cite{yelmen2023deep}, but this line of work has yet to model genome-wide unphased dosage data. 

Hidden Chow-Liu tree (HCLT) models \cite{dang2023HCLTs} provide a tractable probabilistic alternative to neural generators. HCLTs introduce latent variables and learn tree-structured dependencies among genomic variables, often represented as probabilistic circuits to enable efficient likelihood evaluation and sampling. HCLTs provide more explicit probabilistic structure and tractable inference than GANs for example, and their hidden variables increase expressivity over simple tree models to capture longer-range dependencies. In practice, however, the available HCLT-based synthetic genome benchmarks are also tied to restricted SNP panels and do not directly address genome-wide conditional generation from unphased genomes.

Conditional control remains a key gap in synthetic genotype generation. Reference-panel simulators generate phenotypes after genotype simulation by specifying effect sizes, environmental noise, or covariate effects (namely, the objects we are typically interested in learning), and HAPNEST supports large-scale synthetic genotype--phenotype datasets for PRS benchmarking \cite{wharrie2023hapnest}. However, this is not the same as learning a conditional genotype distribution as the genotype is still generated from phased reference haplotype structure, and phenotypes are simply assigned downstream from the generative model. Direct control over genome-wide genotype generation under a target disease-liability is what HELICS offers to address the conditional genome generation gap.

Recent diffusion-based genotype generators move toward this goal but make restrictive design choices. SNPgen \cite{lampis2026snpgen} proposes a phenotype-supervised latent diffusion framework conditioned on binary disease labels, but first precedes generation with GWAS-guided variant selection and models only \(1{,}024\)--\(2{,}048\) trait-associated SNPs with a latent diffusion model. This limits applicability to highly polygenic traits, cross-trait analyses, and discovery settings where the relevant loci are not known in advance. Kenneweg et al. \cite{kenneweg2025generating} also introduce a diffusion model for synthetic human genotypes, but achieve scalability by assigning SNPs to gene regions and using PCA within those regions, which may discard fine-scale nonlinear LD, weak polygenic effects outside selected loci, and long-range cross-chromosomal structure. Their trained models and generated cohorts needed for matched harmonized benchmarking are not released.

\section{Benchmarking generated genome fidelity}
\label{app:benchmarks}

\subsection{Baseline acquisition and generation}

We evaluated baselines using released synthetic data when available and otherwise generated samples from the published reproducible simulation procedures. For the GAN- and RBM-based methods \cite{yelmenGAN}, released synthetic haplotypes were available on the chromosome 15 panel. Ancestry was inferred by matching synthetic haplotypes to 1KG reference haplotypes; EUR-assigned haplotypes were retained, and haplotypes were paired within ancestry to construct diploid dosage matrices. We applied the same procedure to the HCLT-based samples \cite{dang2023HCLTs}, which were generated on the same reference panel. For HAPNEST \cite{wharrie2023hapnest}, we used the released synthetic dosage data directly and retained EUR-assigned samples after ancestry matching to 1KG. For HAPGEN2 \cite{su2011hapgen2} and Sim1000G \cite{dimitromanolakis2019sim1000g}, we simulated synthetic individuals based on phased 1KG EUR haplotypes. Since Sim1000G is computationally expensive at larger variant counts, we simulated only a \(10{,}000\)-variant subset of chromosome 15. For every method, output variants were harmonized using the same build, allele, and strand-matching procedure described in Appendix~\ref{app:variant-harmonization}.

\subsection{Training scale and reference-panel sensitivity}
\label{app:scale_reference}

The available genotype generation baselines compared to HELICS in this study differ in training size, reference cohort source, and computational requirements. For example, the regional learned generators were trained on the 1KG panel, whose Phase 3 release contains only \(2{,}504\) individuals, whereas HELICS supports biobank-scale training. Here, we assess the sensitivity of HELICS to training size and the sensitivity of regional reference-based generators to reference panel source.

To limit computational cost, we retrained chromosome 22 autoencoders and flow models on strictly nested UKB-EUR subsets of \(2{,}500\), \(5{,}000\), \(10{,}000\), and \(100{,}000\) individuals, holding architecture, variants, preprocessing, and external CCDG-EUR evaluation fixed. Subsets were drawn using one permutation of the \(251{,}465\)-individual training pool of the full HELICS-EUR model. The flow architecture used 17 USiT layers, hidden width 512, and 8 attention heads; remaining configurations were kept consistent with the full HELICS model (Table~\ref{tab:flow_training}). Flow training uses batch size 256, stopping after 200 epochs or \(10{,}000\) updates (whichever occurs first). After training, \(1{,}024\) samples were generated from each model. Across the controlled 2.5K to 100K range, common-variant MAF and LD change modestly; overall MAF \(R^2\) remains high across training sizes, with the \(100{,}000\)-individual model achieving the strongest overall and low-frequency MAF agreement (Table~\ref{tab:training_size}). The full HELICS-EUR model is included as a reference (evaluated here on chromosome 22 only), though it differs in both genomic scope and training duration. 

\begin{table}[h]
\centering
\caption{Training-size sensitivity under external CCDG-EUR evaluation.}
\label{tab:training_size}
\small
\begin{tabular}{cccccc}
\toprule
\textbf{Training \(N\)} & \textbf{Training chr.} & \textbf{MAF} \(R^2 \uparrow\) & \textbf{Low-MAF} \(R^2 \uparrow\) & \textbf{MAF MAE} \(\downarrow\) & \textbf{LD diff.} \(\downarrow\) \\
\midrule
2{,}500 & 22 & 0.978 & 0.187 & 0.0199 & 0.205 \\
5{,}000 & 22 & 0.975 & 0.169 & 0.0220 & 0.221 \\
10{,}000 & 22 & 0.968 & 0.179 & 0.0284 & 0.226 \\
100{,}000 & 22 & 0.983 & 0.368 & 0.0173 & 0.210 \\
\midrule
251{,}465 & 1-22 & 0.991 & 0.678 & 0.018 & 0.120 \\
\bottomrule
\end{tabular}
\end{table}

Using the same chromosome 22 setup, we additionally varied flow-model capacity while holding the training data, autoencoder, training schedule, variants, and evaluation fixed (Table~\ref{tab:flow_capacity}). Increasing capacity improves MAF and LD fidelity, and we find that additional training steps further reduce LD discrepancy. Within these tested settings, flow-model capacity has a larger effect on MAF fidelity than increasing training-set size.

\begin{table}[h]
\centering
\caption{Flow-model capacity sensitivity on chromosome 22 under external CCDG-EUR evaluation. The autoencoder, training data ($N = 100{,}000$), and training schedule are fixed. Parameter counts refer to the flow model.}
\label{tab:flow_capacity}
\small
\begin{tabular}{lcccc}
\toprule
\shortstack{\textbf{Flow model}\\\textbf{(parameter count)}}
& \textbf{MAF} \(R^2 \uparrow\)
& \textbf{Low-MAF} \(R^2 \uparrow\)
& \textbf{MAF MAE} \(\downarrow\)
& \textbf{LD diff.} \(\downarrow\) \\
\midrule
Tiny (1.18M) & 0.221 & 0.002 & 0.187 & 0.251 \\
Small (7.98M) & 0.668 & 0.013 & 0.112 & 0.257 \\
Medium (25.5M) & 0.978 & 0.219 & 0.027 & 0.249 \\
Large (58.9M) & 0.983 & 0.260 & 0.024 & 0.248 \\
\bottomrule
\end{tabular}
\end{table}

Next, to assess reference-panel sensitivity, we reran Sim1000G using phased haplotypes from unrelated UKB-EUR individuals instead of 1KG-EUR, holding reference sample sizes and regional variant counts fixed and retaining external CCDG-EUR evaluation. UKB-EUR modestly improves all reported metrics, particularly low-frequency fidelity (Table~\ref{tab:reference_sensitivity}), indicating that reference-panel source affects benchmark performance. HELICS learns directly from biobank-scale unphased dosages, avoiding the need for a phased reference panel.

\begin{table}[h]
\centering
\caption{Sim1000G reference-panel sensitivity against CCDG-EUR. Values are means \(\pm\) standard deviation across 22 independently simulated regions, with one region sampled from each autosome and \(4{,}682\) evaluated variants per region on average.}
\label{tab:reference_sensitivity}
\small
\begin{tabular}{lcccc}
\toprule
Reference & MAF \(R^2\) & MAF MAE & Low-MAF \(R^2\) & LD diff. \\
\midrule
1000G-EUR & \(0.985\pm0.001\) & \(0.012\pm0.0004\) & \(0.579\pm0.034\) & \(0.127\pm0.026\) \\
UKB-EUR & \(0.990\pm0.001\) & \(0.010\pm0.0004\) & \(0.691\pm0.050\) & \(0.122\pm0.026\) \\
\bottomrule
\end{tabular}
\end{table}

\subsection{Genetic data fidelity metrics}
\label{app:benchmark-metric-definitions}

To assess generated data fidelity, we evaluate several properties of the synthetic genomes which are motivated by downstream utility, such as marginal variation, local correlation, ancestry structure, memorization risk, and within-synthetic diversity. The following metrics are assessed in Table~\ref{tab:main_benchmark}:
\begin{itemize}
    \item \textbf{Allele-frequency agreement.}
    For each harmonized variant \(j\), we compute the minor allele frequency (MAF) in the generated cohort and in the reference cohort and report the \(R^2\) between generated and reference MAFs, the mean absolute error (MAE)
    \[
    \mathrm{MAE}_{\mathrm{MAF}}
    =
    \frac{1}{m}\sum_{j=1}^{m}
    \left|
    \widehat f^{\mathrm{gen}}_j-\widehat f^{\mathrm{ref}}_j
    \right|,
    \]
    and a low-frequency \(R^2\) restricted to variants with \(0.01<\mathrm{MAF}<0.05\) in the reference cohort. 
    
    \item \textbf{LD fidelity.}
    We measure correspondence between the generated and real genomes' local SNP correlation structure. For each LD block or fixed genomic window \(b\), let
    \(\boldsymbol R_b^{\mathrm{ref}}\) and \(\boldsymbol R_b^{\mathrm{gen}}\) denote the reference and generated dosage correlation matrices over the variants in that block. We summarize LD distortion using a length-normalized Frobenius discrepancy,
    \[
    d_{\mathrm{LD}}(b)
    =
    \frac{
    \left\|
    \boldsymbol R_b^{\mathrm{gen}}
    -
    \boldsymbol R_b^{\mathrm{ref}}
    \right\|_F
    }{|b|},
    \]
    and report the average across inferred LD blocks. Smaller values indicate closer preservation of local LD. LD blocks were inferred using PLINK with \texttt{---blocks no-pheno-req no-small-max-span}, \texttt{---blocks-inform-frac 0.9}, \texttt{---blocks-max-kb 100000}, \texttt{---blocks-min-maf 0.01}, \texttt{---blocks-recomb-highci 0.6}, \texttt{---blocks-strong-highci 0.8301}, and \texttt{---blocks-strong-lowci 0.5001}. Blocks with insufficient variant overlap after harmonization ($<20$ SNPs) are excluded.

    \item \textbf{Population structure.}
    We assess whether generated samples occupy the same ancestry geometry as real samples. We fit genotype PCs on a reference cohort, standardizing variants using the reference mean and variance, and project generated samples into this fixed PC basis. Agreement is evaluated qualitatively by comparing the generated and reference distributions in PC space. 

    \item \textbf{Membership signal.}
    We use nearest-neighbor IBS similarity to the generated cohort as an empirical memorization proxy. For each real individual \(i\), either from the training set or a held-out validation set, we compute its maximum IBS similarity to any generated individual. We define this distance as pairwise global identity-by-state (IBS) similarity. Between individuals \(x\) and \(y\) as
    \[
    S(x,y)
    =
    1-
    \frac{\sum_{m\in\mathcal M_{xy}} |G_{xm}-G_{ym}|}
    {2|\mathcal M_{xy}|},
    \]
    where \(\mathcal M_{xy}\) is the set of variants nonmissing in both individuals. For each query individual \(x\), its nearest-neighbor similarity to a reference cohort \(\mathcal R\) is
    \(
    S_{\mathrm{NN}}(x,\mathcal R)
    =
    \max_{y\in\mathcal R} S(x,y).
    \)
    Here, \(\mathcal R\) is the generated cohort and the query individuals are either training individuals or held-out validation individuals. We then test whether \(S_{\mathrm{NN}}(x,\mathcal R)\) distinguishes training from validation samples using ROC AUC, reporting \(|\mathrm{AUC}-0.5|\),
    where values near zero indicate that training and validation individuals are similarly close to the generated cohort, while larger values indicate stronger train-validation separability and possible memorization signal. This is not a formal privacy guarantee, but provides a matched empirical check for overfitting.

    \item \textbf{Generated nearest-neighbor IBS similarity.}
    We measure within-synthetic diversity using nearest-neighbor IBS similarity among generated samples.  For each generated individual, we compute its maximum IBS similarity to another generated individual and summarize the distribution. From the definitions above, \(\mathcal R\) here is the generated cohort itself and the self-match is excluded, so high \(S_{\mathrm{NN}}\) indicates near-duplicate generated samples or mode collapse. Lower values indicate greater generated diversity.
\end{itemize}

In addition to the nearest-neighbor IBS metrics above, we infer kinship using KING. While IBS nearest-neighbor similarity measures overall genotype similarity by allele sharing across matched variants, KING estimates kinship coefficients which are intended to reflect identity-by-descent-like relatedness categories; this provides an interpretable degree-of-relatedness summary. We apply KING in three settings. We first compare synthetic samples to the corresponding training reference panel to assess train proximity. For HELICS, the training reference is UKB, while for Sim1000G, HAPGEN2, and HAPNEST, the training reference is the 1KG cohort. We then compare held-out validation samples to the same training reference, providing a matched baseline for relatedness under each method's harmonized variant set. We also estimate within-synthetic kinship to assess generated diversity; excess close relatedness among generated samples indicates duplication or mode collapse.

Note that these results should be interpreted relatively rather than as calibrated genome-wide relatedness estimates. We therefore exclude small regional generators with substantially fewer variants, where kinship inference is less reliable and often inflated. On this relative chromosome 15 benchmark, HELICS produces no MZ-twin, first-degree, or second-degree synthetic-to-training relationships, and its third-degree synthetic-to-training count is below the matched validation-to-training baseline. Within the generated cohort, HELICS also shows no close within-synthetic relationships and fewer third-degree pairs than Sim1000G or HAPGEN2, supporting the IBS-based diversity analysis (Table~\ref{tab:king_relatedness}).

\clearpage 
\section{Additional results}
\label{app:additional-res}

\begin{table}[h]
\centering
\caption{
KING kinship analysis on the chromosome 15 benchmark panel. Genetic relatedness is estimated from genotype data, and pairs are categorized by the inferred degree of relatedness. The validation-to-training panel provides a matched baseline for each method's variant overlap set. Validation samples are unrelated held-out reference individuals compared against the same training reference used for each method. Fewer close relatives and more unrelated pairs indicate lower train proximity.
}
\label{tab:king_relatedness}
\begin{tabular}{lrrrr}
\toprule
\multicolumn{5}{c}{\textbf{Synthetic samples vs. training reference}} \\
\midrule
Relatedness category & Sim1000G & HELICS & HAPGEN2 & HAPNEST \\
\midrule
MZ twins \(\downarrow\)      & 0 & 0 & 0 & 0 \\
1st degree \(\downarrow\)    & 0 & 0 & 0 & 0 \\
2nd degree \(\downarrow\)    & 0 & 0 & 0 & 0 \\
3rd degree \(\downarrow\)    & 15{,}656 & 21{,}645 & 1{,}779 & 29{,}114 \\
Unrelated \(\uparrow\)       & 4{,}984{,}344 & 24{,}978{,}355 & 24{,}998{,}221 & 24{,}970{,}886 \\
Total pairs                  & 5{,}000{,}000 & 25{,}000{,}000 & 25{,}000{,}000 & 25{,}000{,}000 \\
\midrule
\multicolumn{5}{c}{\textbf{Validation samples vs. training reference}} \\
\midrule
Relatedness category & Sim1000G & HELICS & HAPGEN2 & HAPNEST \\
\midrule
MZ twins \(\downarrow\)      & 0 & 0 & 0 & 0 \\
1st degree \(\downarrow\)    & 0 & 0 & 0 & 0 \\
2nd degree \(\downarrow\)    & 3 & 3 & 2 & 2 \\
3rd degree \(\downarrow\)    & 103{,}748 & 61{,}147 & 67{,}137 & 94{,}064 \\
Unrelated \(\uparrow\)       & 24{,}896{,}249 & 24{,}938{,}850 & 24{,}932{,}861 & 24{,}905{,}934 \\
Total pairs                  & 25{,}000{,}000 & 25{,}000{,}000 & 25{,}000{,}000 & 25{,}000{,}000 \\
\midrule
\multicolumn{5}{c}{\textbf{Within synthetic cohort}} \\
\midrule
Relatedness category & Sim1000G & HELICS & HAPGEN2 & HAPNEST \\
\midrule
MZ twins \(\downarrow\)      & 0 & 0 & 0 & 0 \\
1st degree \(\downarrow\)    & 0 & 0 & 37 & 0 \\
2nd degree \(\downarrow\)    & 0 & 0 & 161 & 0 \\
3rd degree \(\downarrow\)    & 3{,}603 & 769 & 2{,}361 & 102 \\
Unrelated \(\uparrow\)       & 495{,}897 & 498{,}731 & 496{,}941 & 499{,}398 \\
Total pairs                  & 499{,}500 & 499{,}500 & 499{,}500 & 499{,}500 \\
\bottomrule
\end{tabular}
\end{table}

\begin{table}[t]
\centering
\caption{
Chromosome-specific fidelity metrics for HELICS-EUR based on 5{,}000 generated samples. Metrics are computed separately for each autosome using harmonized UKB training and generated variants. MAF \(R^2\) and MAF MAE compare generated and UKB training minor allele frequencies across all variants on the chromosome. Low-MAF \(R^2\) is the \(R^2\) restricted to low-frequency variants defined by UKB MAF \(<0.05\); MAF \(<0.2\) \(R^2\) is the corresponding \(R^2\) for variants with UKB MAF \(<0.2\). LD diff. is the mean normalized Frobenius norm of the discrepancy between UKB and generated SNP--SNP LD correlation matrices within predefined LD blocks; lower values indicate better LD fidelity. LD blocks denotes the number of evaluated blocks on each chromosome.
}
\label{tab:chromosome_specific_fidelity_helics_eur}
\small
\setlength{\tabcolsep}{4pt}
\renewcommand{\arraystretch}{0.95}
\begin{tabular}{c r r r r r r}
\toprule
\textbf{Chr}
& \textbf{MAF \(R^2\) \(\uparrow\)}
& \textbf{Low-MAF \(R^2\) \(\uparrow\)}
& \textbf{MAF \(<0.2\) \(R^2\) \(\uparrow\)}
& \textbf{MAF MAE \(\downarrow\)}
& \textbf{LD diff. \(\downarrow\)}
& \textbf{LD blocks} \\
\midrule
1  & 0.9805 & 0.4967 & 0.9205 & 0.0248 & 0.1414 & 753 \\
2  & 0.9735 & 0.5144 & 0.9037 & 0.0229 & 0.1452 & 808 \\
3  & 0.9765 & 0.5307 & 0.9117 & 0.0269 & 0.1393 & 639 \\
4  & 0.9774 & 0.5203 & 0.9106 & 0.0258 & 0.1327 & 631 \\
5  & 0.9739 & 0.4972 & 0.8955 & 0.0262 & 0.1327 & 597 \\
6  & 0.9793 & 0.6279 & 0.9197 & 0.0261 & 0.1362 & 636 \\
7  & 0.9812 & 0.6592 & 0.9228 & 0.0254 & 0.1301 & 534 \\
8  & 0.9789 & 0.5733 & 0.9193 & 0.0233 & 0.1325 & 512 \\
9  & 0.9908 & 0.6593 & 0.9664 & 0.0151 & 0.1301 & 378 \\
10 & 0.9755 & 0.3983 & 0.9088 & 0.0258 & 0.1462 & 480 \\
11 & 0.9792 & 0.5473 & 0.9304 & 0.0261 & 0.1551 & 463 \\
12 & 0.9760 & 0.4345 & 0.9112 & 0.0241 & 0.1481 & 435 \\
13 & 0.9805 & 0.4190 & 0.9367 & 0.0271 & 0.1482 & 330 \\
14 & 0.9745 & 0.5801 & 0.9204 & 0.0182 & 0.0816 & 286 \\
15 & 0.9646 & 0.5055 & 0.8995 & 0.0199 & 0.1022 & 248 \\
16 & 0.9846 & 0.5696 & 0.9479 & 0.0220 & 0.1352 & 293 \\
17 & 0.9589 & 0.5459 & 0.9078 & 0.0216 & 0.1151 & 249 \\
18 & 0.9573 & 0.3312 & 0.8731 & 0.0214 & 0.0937 & 255 \\
19 & 0.9771 & 0.4680 & 0.9314 & 0.0265 & 0.1551 & 218 \\
20 & 0.9842 & 0.5584 & 0.9343 & 0.0261 & 0.1392 & 217 \\
21 & 0.9683 & 0.5167 & 0.8896 & 0.0205 & 0.0888 & 124 \\
22 & 0.9912 & 0.6775 & 0.9593 & 0.0180 & 0.1264 & 109 \\
\bottomrule
\end{tabular}
\end{table}

\begin{table}[t]
\centering
\caption{
Chromosome-specific fidelity metrics for HELICS based on 5{,}000 generated samples. Metrics are computed separately for each autosome using harmonized UKB training and generated variants. MAF \(R^2\) and MAF MAE compare generated and UKB training minor allele frequencies across all variants on the chromosome. Low-MAF \(R^2\) is the \(R^2\) restricted to low-frequency variants defined by UKB MAF \(<0.05\); MAF \(<0.2\) \(R^2\) is the corresponding \(R^2\) for variants with UKB MAF \(<0.2\). LD diff. is the mean normalized Frobenius norm of the discrepancy between UKB and generated SNP--SNP LD correlation matrices within predefined LD blocks; lower values indicate better LD fidelity. LD blocks denotes the number of evaluated blocks on each chromosome.
}
\label{tab:chromosome_specific_fidelity_helics}
\small
\setlength{\tabcolsep}{4pt}
\renewcommand{\arraystretch}{0.95}
\begin{tabular}{c r r r r r r}
\toprule
\textbf{Chr}
& \textbf{MAF \(R^2\) \(\uparrow\)}
& \textbf{Low-MAF \(R^2\) \(\uparrow\)}
& \textbf{MAF \(<0.2\) \(R^2\) \(\uparrow\)}
& \textbf{MAF MAE \(\downarrow\)}
& \textbf{LD diff. \(\downarrow\)}
& \textbf{LD blocks} \\
\midrule
1  & 0.9790 & 0.2032 & 0.8503 & 0.0330 & 0.1616 & 753 \\
2  & 0.9876 & 0.5716 & 0.9566 & 0.0179 & 0.1082 & 808 \\
3  & 0.9871 & 0.8617 & 0.9830 & 0.0170 & 0.0908 & 639 \\
4  & 0.9750 & 0.7342 & 0.9585 & 0.0232 & 0.0909 & 631 \\
5  & 0.9886 & 0.3162 & 0.9159 & 0.0240 & 0.1428 & 597 \\
6  & 0.9868 & 0.7083 & 0.9546 & 0.0168 & 0.1113 & 636 \\
7  & 0.9941 & 0.7007 & 0.9746 & 0.0160 & 0.1074 & 534 \\
8  & 0.9740 & 0.7176 & 0.9574 & 0.0240 & 0.1001 & 512 \\
9  & 0.9920 & 0.5687 & 0.9575 & 0.0184 & 0.1170 & 378 \\
10 & 0.9873 & 0.4019 & 0.9141 & 0.0269 & 0.1391 & 480 \\
11 & 0.9835 & 0.6799 & 0.9567 & 0.0188 & 0.1085 & 463 \\
12 & 0.9761 & 0.7211 & 0.9606 & 0.0225 & 0.1025 & 435 \\
13 & 0.9940 & 0.6962 & 0.9783 & 0.0140 & 0.1024 & 330 \\
14 & 0.9888 & 0.7132 & 0.9777 & 0.0158 & 0.0929 & 286 \\
15 & 0.9901 & 0.3270 & 0.9323 & 0.0227 & 0.1258 & 248 \\
16 & 0.9906 & 0.4326 & 0.9323 & 0.0233 & 0.1259 & 293 \\
17 & 0.9638 & 0.0703 & 0.7831 & 0.0413 & 0.1881 & 249 \\
18 & 0.9806 & 0.7869 & 0.9732 & 0.0224 & 0.0815 & 255 \\
19 & 0.9854 & 0.6829 & 0.9679 & 0.0185 & 0.1204 & 218 \\
20 & 0.9927 & 0.7391 & 0.9505 & 0.0230 & 0.1154 & 217 \\
21 & 0.9808 & 0.4225 & 0.9426 & 0.0183 & 0.1087 & 124 \\
22 & 0.9869 & 0.3356 & 0.9363 & 0.0174 & 0.1229 & 109 \\
\bottomrule
\end{tabular}
\end{table}

\begin{table}[t]
\centering
\caption{
Ancestry-specific genotype fidelity of multi-ancestry HELICS across all 22 autosomes. \(N_{\mathrm{train}}\) denotes the number of training individuals in each ancestry group: European (EUR), South Asian (SAS), East Asian (EAS), and African (AFR). Values are means \(\pm\) standard deviations across chromosomes. Allele-frequency fidelity is weaker in underrepresented groups, particularly for low-frequency variants.}
\label{tab:ancestry_fidelity}
\small
\setlength{\tabcolsep}{4pt}
\begin{tabular}{lrcccc}
\toprule
\textbf{Ancestry}
& \(N_{\mathrm{train}}\)
& \textbf{MAF} \(R^2 \uparrow\)
& \textbf{MAF MAE} \(\downarrow\)
& \textbf{Low-MAF} \(R^2 \uparrow\)
& \textbf{LD diff.} \(\downarrow\) \\
\midrule
EUR & 391{,}125 & \(0.980\pm0.010\) & \(0.023\pm0.006\) & \(0.488\pm0.199\) & \(0.109\pm0.013\) \\
SAS & 9{,}033   & \(0.807\pm0.067\) & \(0.046\pm0.007\) & \(0.080\pm0.026\) & \(0.112\pm0.011\) \\
EAS & 5{,}576   & \(0.718\pm0.068\) & \(0.049\pm0.006\) & \(0.008\pm0.006\) & \(0.117\pm0.008\) \\
AFR & 7{,}291   & \(0.710\pm0.084\) & \(0.051\pm0.007\) & \(0.069\pm0.054\) & \(0.116\pm0.017\) \\
\bottomrule
\end{tabular}
\end{table}

\begin{figure}[t]
    \centering
    \includegraphics[width=\linewidth]{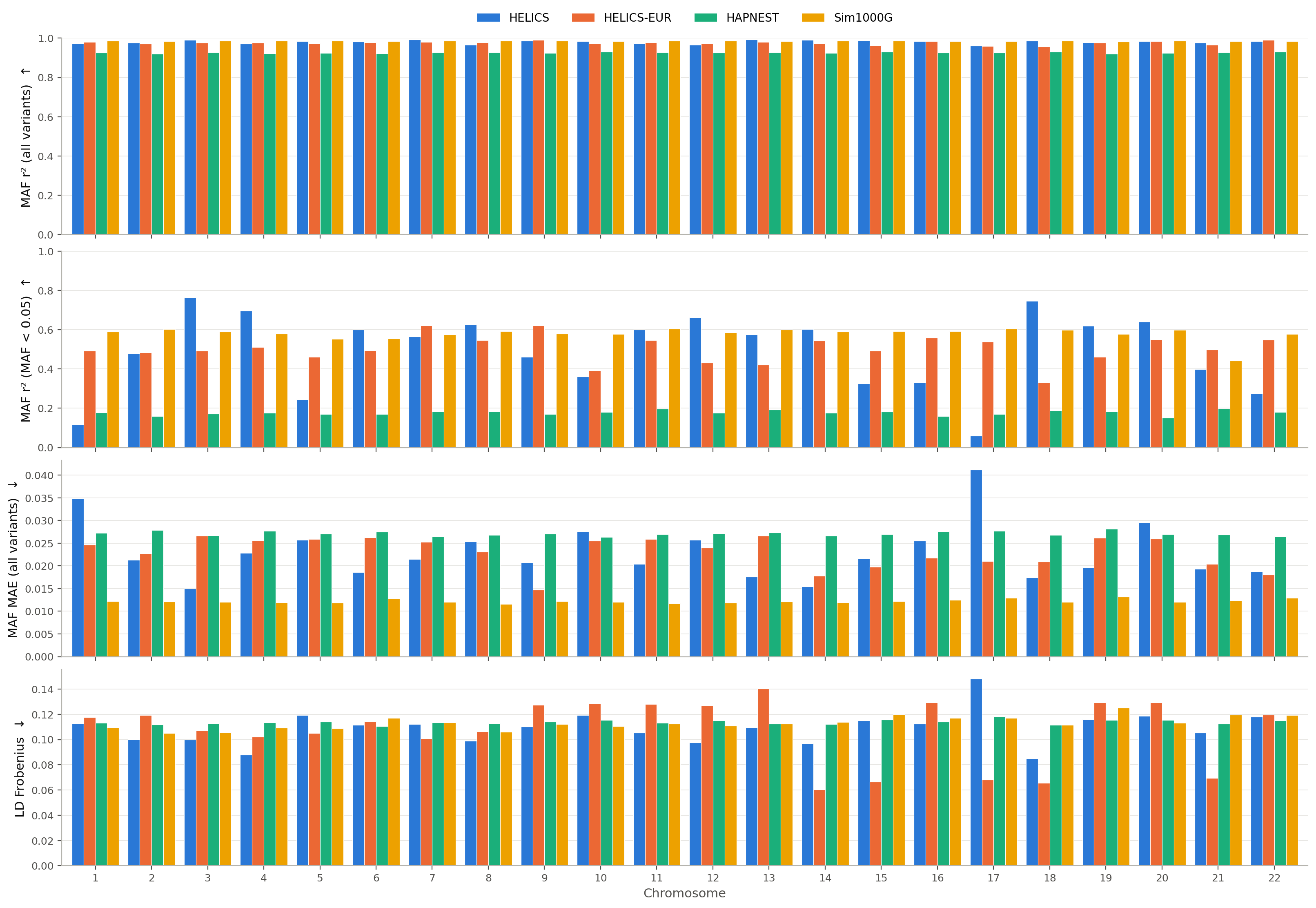}
    \caption{Chromosome-resolved EUR genotype fidelity against the external CCDG-EUR cohort across all 22 autosomes. Rows show (top to bottom) overall MAF \(R^2\), low-frequency MAF \(R^2\) (\(0.01<\mathrm{MAF}<0.05\)), MAF MAE, and mean normalized Frobenius LD discrepancy. HELICS, HELICS-EUR, and HAPNEST generate genome-wide samples; Sim1000G is evaluated on one independently simulated region per chromosome. Each method uses its available overlap with the harmonized reference panel. Higher \(R^2\) and lower MAE and LD discrepancy indicate closer agreement.}
    \label{fig:ancestry_fidelity_chromosomes}
\end{figure}

\begin{table}[h]
\centering
\caption{
Within-ancestry alignment between real and HELICS-generated genome-wide latent representations in the multi-ancestry model. \(N_{\mathrm{train}}\) denotes the number of training individuals from each ancestry group used to fit HELICS. For each ancestry, we compare real and generated latent distributions using sliced Wasserstein distance and RBF MMD\(^2\), where smaller values indicate closer agreement. Means and standard errors are estimated from 250 bootstrap replicates, each formed by sampling 542 real and 542 generated latents within ancestry. The results show very close agreement for the majority White ancestry group, which is also the largest training group, and systematically weaker alignment for underrepresented ancestry groups. This pattern is consistent with the reduced representation of these groups during training and suggests that latent-space fidelity tracks training sample size in the multi-ancestry setting.
}
\label{app:tab:latent_alignment}
\begin{tabular}{lrcc}
\toprule
Ancestry & \(N_{\mathrm{train}}\) & Sliced \(W_2\) & RBF MMD\(^2\) \\
\midrule
White & 391{,}125 & 0.002 $\pm$ 0.0001 & 0.002 $\pm$ 0.0003 \\
South Asian & 9{,}033 & 0.042 $\pm$ 0.001 & 0.315 $\pm$ 0.016 \\
SE Asian & 5{,}576 & 0.078 $\pm$ 0.003 & 0.331 $\pm$ 0.016 \\
Black & 4{,}230 & 0.075 $\pm$ 0.002 & 0.846 $\pm$ 0.037 \\
African & 3{,}061 & 0.085 $\pm$ 0.003 & 0.564 $\pm$ 0.016 \\
\bottomrule
\end{tabular}
\end{table}

\begin{figure}[h]
    \centering
    \includegraphics[width=\linewidth]{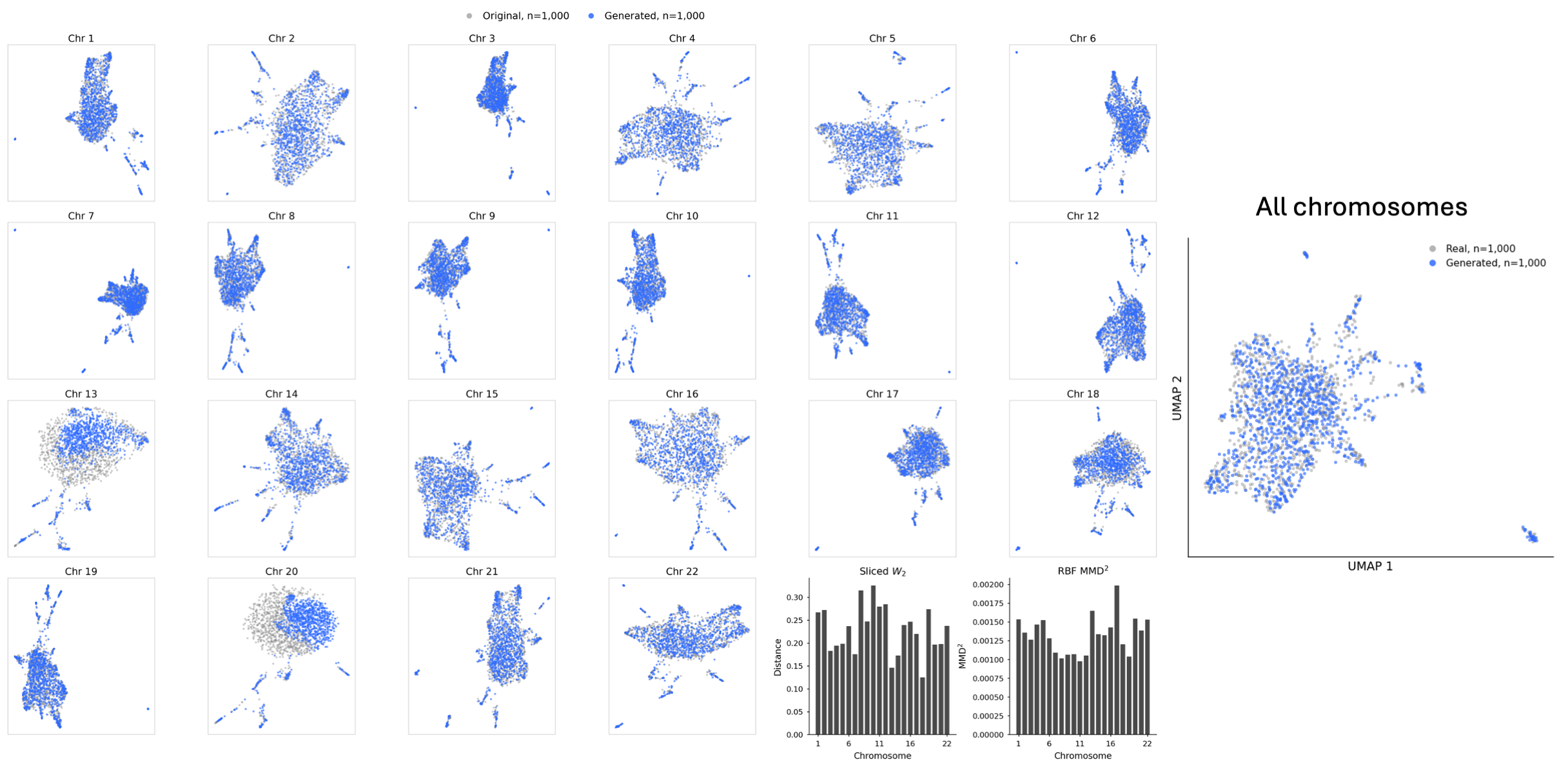}
    \caption{
    UMAP visualization of HELICS latent representations for real validation individuals and generated individuals, shown separately for each chromosome and for the concatenated genome-wide latent space. Each chromosome-specific panel shows a two-dimensional UMAP embedding of the corresponding chromosome latent tokens for \(1{,}000\) held-out real individuals and \(1{,}000\) generated individuals. The large panel at right shows the same comparison using the concatenated latent representation across all chromosomes. Across nearly all chromosomes, generated latents overlap closely with real latents and do not form a clearly separated cluster, suggesting that the latent flow model covers the support of the chromosome-specific latent spaces rather than generating isolated or low-diversity samples. The bar plots summarize chromosome-wise sliced Wasserstein distance and RBF MMD\(^2\) between real and generated latent distributions; lower values indicate better alignment. These quantitative summaries mirror the visual UMAP overlap and show that latent fidelity is broadly consistent across chromosomes.
    }
    \label{fig:chrom-UMAPs}
\end{figure}


\begin{figure}[h]
    \centering
    \includegraphics[width=\linewidth]{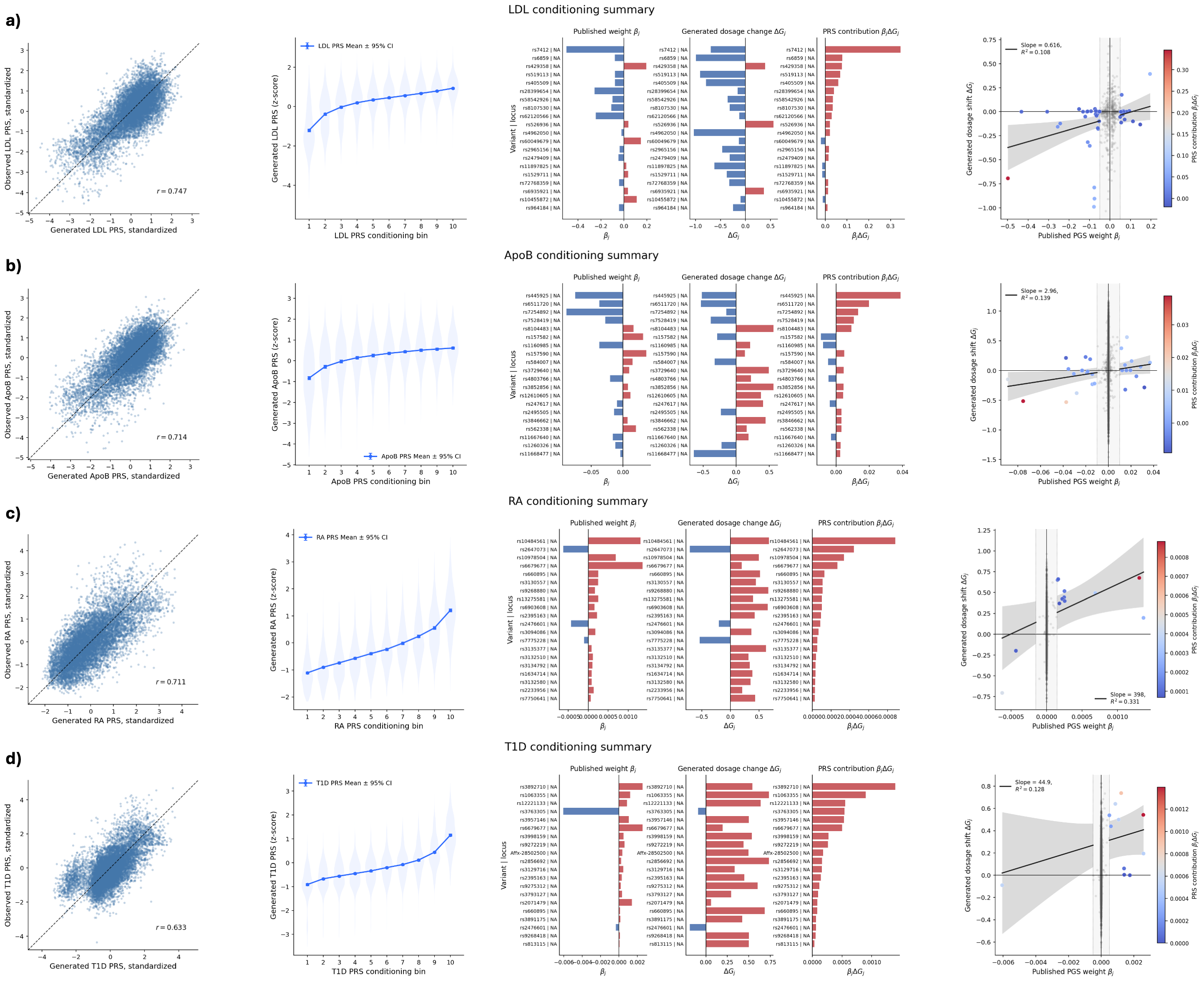}
    \caption{
    Trait-guided perturbation analyses for low-density lipoprotein (LDL; a), apolipoprotein B (ApoB; b), rheumatoid arthritis (RA; c), and type 1 diabetes (T1D; d). 
    Each row repeats the same analysis used for HDL in Figure~\ref{fig:modelInterp}. 
    Left: PRS computed for generated genomes conditioned on validation sample conditioning profiles is compared to the observed PRS of the corresponding real validation individual; the reported \(r\) measures conditional control success. 
    Second: the target PRS conditioning bin is swept from 1 to 10 while ancestry and all other trait conditions are fixed; violins show the distribution of generated PRS and points show the mean with 95\% confidence intervals. 
    Third: the top 20 variants with the largest absolute PRS-dosage shifts contributions (\(\beta_j\Delta G_j\)) are decomposed into the published PRS weight \(\beta_j\), the generated effect-allele dosage shift \(\Delta G_j=\mathbb E[\widetilde G_j(10)]-\mathbb E[\widetilde G_j(1)]\). 
    Right: across matched PRS variants, generated dosage shifts are compared to published PRS weights; points outside the shaded low-effect region are used for the displayed linear fit, and color denotes \(\beta_j\Delta G_j\). 
    Across all traits, HELICS recovers individual-level PRS variation, produces monotone low-to-high PRS shifts under conditioning-bin perturbations, and induces SNP-level dosage changes that are broadly concordant with large published association directions, despite being trained only on discretized PRS summaries rather than variant-level PRS weights.
    }
    \label{fig:modelInterp-othertraits}
\end{figure}

\begin{table}[t]
\centering
\caption{Ancestry-stratified conditional-control results from multi-ancestry HELICS for European (EUR), South Asian (SAS), and African (AFR) ancestry groups. The perturbation analysis partitions the same \(1{,}001\) validation profiles used in Table~\ref{tab:conditional_control}. \(N_{\mathrm{corr}}\) denotes the number of paired real and generated genomes used to compute the Pearson correlation between continuous PRSs under validation-profile conditioning (Gen.-obs. \(r\)); \(N_{\mathrm{sweep}}\) denotes the number of profiles used for each perturbation condition. Perturbations change only the target trait's PRS decile, holding ancestry and other trait conditions fixed. Bin 10 \(>\) bin 1 is the proportion of profiles whose generated PRS under bin-10 conditioning exceeds that under bin-1 conditioning for the same profile. Top-20 sign fraction is the unweighted fraction of the 20 variants with the largest absolute PRS contributions \(|\beta_j\Delta G_j|\) satisfying \(\beta_j\Delta G_j>0\), with dosage shifts and rankings computed separately within each ancestry. Low-to-high PRS responses and SNP-level directional alignment remain broadly preserved across groups, while generated-observed PRS correlations are lower in SAS and AFR.}
\label{tab:conditional_control_ancestry}
\small
\setlength{\tabcolsep}{3pt}
\renewcommand{\arraystretch}{1.05}
\begin{tabular}{llrrrccc}
\toprule
Trait & Ancestry
& \(N_{\mathrm{corr}}\)
& \(N_{\mathrm{sweep}}\)
& \shortstack{Matched\\PGS variants}
& \shortstack{Gen.-obs.\\\(r\)}
& \shortstack{Bin 10 \(>\)\\bin 1 mean}
& \shortstack{Top-20\\alignment} \\
\midrule
HDL & EUR & 9{,}366 & 950 & 168 & 0.684 & 0.977 & 0.950 \\
HDL & SAS & 213 & 28 & 168 & 0.230 & 1.000 & 0.950 \\
HDL & AFR & 172 & 23 & 168 & 0.244 & 0.913 & 0.900 \\
\midrule
LDL & EUR & 9{,}366 & 950 & 606 & 0.762 & 0.997 & 0.800 \\
LDL & SAS & 213 & 28 & 606 & 0.221 & 1.000 & 0.750 \\
LDL & AFR & 172 & 23 & 606 & 0.308 & 0.957 & 0.850 \\
\midrule
T1D & EUR & 9{,}366 & 950 & 16{,}570 & 0.648 & 0.998 & 1.000 \\
T1D & SAS & 213 & 28 & 16{,}570 & 0.408 & 1.000 & 1.000 \\
T1D & AFR & 172 & 23 & 16{,}570 & 0.031 & 1.000 & 1.000 \\
\midrule
RA & EUR & 9{,}366 & 950 & 14{,}728 & 0.724 & 0.991 & 1.000 \\
RA & SAS & 213 & 28 & 14{,}728 & 0.163 & 1.000 & 1.000 \\
RA & AFR & 172 & 23 & 14{,}728 & 0.243 & 1.000 & 0.950 \\
\midrule
ApoB & EUR & 9{,}366 & 950 & 43{,}001 & 0.742 & 0.946 & 0.700 \\
ApoB & SAS & 213 & 28 & 43{,}001 & 0.157 & 0.964 & 0.700 \\
ApoB & AFR & 172 & 23 & 43{,}001 & 0.205 & 0.913 & 0.650 \\
\bottomrule
\end{tabular}
\end{table}

\clearpage

\section*{NeurIPS Paper Checklist}

\begin{enumerate}

\item {\bf Claims}
    \item[] Question: Do the main claims made in the abstract and introduction accurately reflect the paper's contributions and scope?
    \item[] Answer: \answerYes{}
    \item[] Justification: The abstract and introduction state the paper's contributions as a scalable latent generative model for unphased genome-wide SNP dosage data, with evaluations of fidelity, diversity, privacy-proxy behavior, ancestry structure, and conditional generation. These claims are supported by the main experiments and are scoped to the evaluated UKB, 1KG, CCDG, and baseline benchmark settings.

\item {\bf Limitations}
    \item[] Question: Does the paper discuss the limitations of the work performed by the authors?
    \item[] Answer: \answerYes{} 
    \item[] Justification: The paper discusses several limitations such as the use of empirical memorization proxies rather than formal privacy guarantees, the need for statistically robust frameworks to leverage synthetic data for downstream inference and discovery, and the restriction to common variants. These limitations are discussed throughout in the main text.

\item {\bf Theory assumptions and proofs}
    \item[] Question: For each theoretical result, does the paper provide the full set of assumptions and a complete (and correct) proof?
    \item[] Answer: \answerNA{} 
    \item[] Justification: The paper does not make formal theoretical claims or present theorems requiring proof. Mathematical expressions are used to define the model, training objectives, and evaluation metrics, with implementation details provided in the methods and appendix.

    \item {\bf Experimental result reproducibility}
    \item[] Question: Does the paper fully disclose all the information needed to reproduce the main experimental results of the paper to the extent that it affects the main claims and/or conclusions of the paper (regardless of whether the code and data are provided or not)?
    \item[] Answer: \answerYes{} 
    \item[] Justification: The paper and appendix specify the data preprocessing, variant harmonization, model architecture, training objectives, hyperparameters, generation procedure, calibration procedure, baselines, and evaluation metrics needed to reproduce the experimental pipeline. Exact reproduction of UKB-trained models requires authorized access to the genotype data, but the methodological details needed to replicate the experiments under this data access are provided in Appendix~\ref{app:data-prep}, Appendix~\ref{app:architecture-details}, Appendix~\ref{app:conditions}, and Appendix~\ref{app:benchmarks}.


\item {\bf Open access to data and code}
    \item[] Question: Does the paper provide open access to the data and code, with sufficient instructions to faithfully reproduce the main experimental results, as described in supplemental material?
    \item[] Answer: \answerNo{} 
    \item[] Justification: Due to access restrictions and participant privacy, individual-level UK Biobank genotypes and derived individual-level genomic data cannot be publicly redistributed. We provide detailed methodological and experimental specifications in the paper and appendix, and release code implementing the HELICS architecture and training pipeline (See Appendix \ref{app:data-prep}); the released code contains the model implementation and enables reproduction by researchers with access to the same datasets.

\item {\bf Experimental setting/details}
    \item[] Question: Does the paper specify all the training and test details (e.g., data splits, hyperparameters, how they were chosen, type of optimizer) necessary to understand the results?
    \item[] Answer: \answerYes{} 
    \item[] Justification: The paper specifies the UKB data preprocessing and data split construction, the HELICS-EUR training subset, external reference cohorts, harmonization procedure (Appendix \ref{app:data-prep}), chromosome-specific autoencoder configuration, latent flow training setup, sampling procedure, and benchmark construction. Hyperparameters for both the autoencoders and latent flow model are reported in Appendix~\ref{app:AEarchitecture-details} and Appendix~\ref{app:usit_details}.

\item {\bf Experiment statistical significance}
    \item[] Question: Does the paper report error bars suitably and correctly defined or other appropriate information about the statistical significance of the experiments?
    \item[] Answer: \answerYes{} 
    \item[] Justification: The paper reports uncertainty estimates for the perturbation analysis (e.g., 95\% confidence intervals around PRS estimates) and trial repetition details for stochastic evaluations, including bootstrap standard errors for latent-distribution comparisons and sample-size information for conditional generation and fidelity analyses. Tables and figures specify the number of replicates, generated samples, evaluation samples, and harmonized variants used.

\item {\bf Experiments compute resources}
    \item[] Question: For each experiment, does the paper provide sufficient information on the computer resources (type of compute workers, memory, time of execution) needed to reproduce the experiments?
    \item[] Answer: \answerYes{} 
    \item[] Justification: The appendix reports the main training hardware and configuration for the latent flow model, such as the use of \(4\times\) NVIDIA H200 GPUs, training steps, sampling steps, and training budget. See Appendix \ref{app:architecture-details}.
    
\item {\bf Code of ethics}
    \item[] Question: Does the research conducted in the paper conform, in every respect, with the NeurIPS Code of Ethics \url{https://neurips.cc/public/EthicsGuidelines}?
    \item[] Answer: \answerYes{} 
    \item[] Justification:  The research fully conforms to the NeurIPS Code of Ethics. Use of UKB data was approved under application 52008. The paper discusses limitations and risks associated with synthetic genomic data, such as privacy and fairness.

\item {\bf Broader impacts}
    \item[] Question: Does the paper discuss both potential positive societal impacts and negative societal impacts of the work performed?
    \item[] Answer: \answerYes{} 
    \item[] Justification: The paper discusses potential positive impacts, such as scalable genomic simulation, benchmark construction, privacy-aware data sharing research, and controlled generation for genetic discovery. Potential risks are also discussed, such as the misuse of synthetic genomes, incomplete privacy protection, and ancestry-representation concerns.
    
\item {\bf Safeguards}
    \item[] Question: Does the paper describe safeguards that have been put in place for responsible release of data or models that have a high risk for misuse (e.g., pre-trained language models, image generators, or scraped datasets)?
    \item[] Answer: \answerYes{} 
    \item[] Justification: The paper does not release restricted individual-level data or trained generative model checkpoints. Reported outputs are aggregate benchmark statistics and figures, and the evaluation includes privacy-proxy and relatedness analyses designed to identify memorization or near-duplicate generation behavior.

\item {\bf Licenses for existing assets}
    \item[] Question: Are the creators or original owners of assets (e.g., code, data, models), used in the paper, properly credited and are the license and terms of use explicitly mentioned and properly respected?
    \item[] Answer: \answerYes{} 
    \item[] Justification: The paper cites the existing datasets, software, and baseline methods used in the study, including UK Biobank, 1000 Genomes, CCDG, PGS Catalog scores, and existing synthetic genome generators. Controlled-access resources were used under their applicable data-use terms, and public software, released synthetic data, and score resources are credited through the corresponding publications and source documentation.

\item {\bf New assets}
    \item[] Question: Are new assets introduced in the paper well documented and is the documentation provided alongside the assets?
    \item[] Answer: \answerNA{} 
    \item[] Justification: The paper does not release a new public dataset, trained model, or benchmark asset. The model, generated data construction, and benchmark procedures are documented in the paper and appendix, but restricted genomic assets are not redistributed.

\item {\bf Crowdsourcing and research with human subjects}
    \item[] Question: For crowdsourcing experiments and research with human subjects, does the paper include the full text of instructions given to participants and screenshots, if applicable, as well as details about compensation (if any)? 
    \item[] Answer: \answerNA{} 
    \item[] Justification: The paper does not involve crowdsourcing, interactive user studies, or human-subject experiments conducted by the authors. 

\item {\bf Institutional review board (IRB) approvals or equivalent for research with human subjects}
    \item[] Question: Does the paper describe potential risks incurred by study participants, whether such risks were disclosed to the subjects, and whether Institutional Review Board (IRB) approvals (or an equivalent approval/review based on the requirements of your country or institution) were obtained?
    \item[] Answer: \answerYes{} 
    \item[] Justification: Use of UKB data was approved under application 52008. Privacy risks and potential misuse of genomic data are mitigated by using controlled-access data and not releasing restricted individual-level genotypes.

\item {\bf Declaration of LLM usage}
    \item[] Question: Does the paper describe the usage of LLMs if it is an important, original, or non-standard component of the core methods in this research? Note that if the LLM is used only for writing, editing, or formatting purposes and does \emph{not} impact the core methodology, scientific rigor, or originality of the research, declaration is not required.
    \item[] Answer: \answerNA{} 
    \item[] Justification: The core methodology does not use LLMs as a model component, experimental baseline, data-generation procedure, or evaluation method. 

\end{enumerate}

\end{document}